\documentclass[11pt,a4paper]{article}
\usepackage[T1]{fontenc}
\usepackage[utf8]{inputenc}
\usepackage{latexsym}
\usepackage{amsmath}
\usepackage{bbm}
\usepackage{amssymb}
\usepackage{babel}
\usepackage{amsfonts}
\usepackage{times}
\usepackage{theorem}
\usepackage{epsfig}
\usepackage{enumerate}
\usepackage{color,xcolor}
\usepackage{xspace,needspace,xparse}
\usepackage[active]{srcltx}
\usepackage[colorlinks=true]{hyperref}
\usepackage{fancyhdr}
\usepackage{booktabs}
\colorlet{darkblue}{blue!50!black}
\hypersetup{
	colorlinks,%
	citecolor=darkblue,%
	filecolor=red,%
	linkcolor=darkblue,%
	urlcolor=darkblue,%
	pdfnewwindow=true,%
	pdfstartview={FitH}
}

\numberwithin{equation}{section}

\newcounter{smallarabics}

\newcounter{smallroman}

\newcommand{\ben}{\begin{enumerate}[{\bf (1)}]}
\newcommand{\een}{\end{enumerate}}

\newtheorem{theorem}{Theorem}[section]
\newtheorem{proposition}[theorem]{Proposition}
\newtheorem{lemma}[theorem]{Lemma}
\newtheorem{definition}[theorem]{Definition}

\usepackage[normalem]{ulem}

\AtBeginEnvironment{theorem}{\Needspace{5\baselineskip}}
\AtBeginEnvironment{proposition}{\Needspace{3\baselineskip}}
\AtBeginEnvironment{definition}{\Needspace{5\baselineskip}}
\AtBeginEnvironment{corollary}{\Needspace{5\baselineskip}}
\AtBeginEnvironment{lemma}{\Needspace{5\baselineskip}}
\AtBeginEnvironment{quote}{\Needspace{5\baselineskip}}

 \def\cH{{\mathcal H}} \def\cI{{\mathcal I}}
 \def\cK{{\mathcal K}} \def\cL{{\mathcal L}}
 \def\cN{{\mathcal N}} \def\cO{{\mathcal O}}
  
\def\cS{{\mathcal S}}  
  
  \def\cNo{{\mathcal N}_0}

\def\R{{\mathbb R}}

\def\rr{{\mathbb R}}

\def\cc{{\mathbb C}}
\def\nn{{\mathbb N}}

\def\slim{\mathop{\mathrm{s\!\!-\!\!lim}}}

\def\tr{{\rm tr}}

\def\e{{\rm e}}
\def\i{{\rm i}}

\def\d{{\rm d}}

\def\bep{\begin{proposition}}
\def\eep{\end{proposition}}
\def\bet{\begin{theoreme}}
\def\eet{\end{theoreme}}
\def\bel{\begin{lemma}}
\def\eel{\end{lemma}}

\renewcommand\Re{{\mathrm{Re}\,}}
\renewcommand\Im{{\mathrm{Im}\,}}

\def\one{{\mathchoice {\rm 1\mskip-4mu l} {\rm 1\mskip-4mu l} {\rm 1\mskip-4.5mu l} {\rm 1\mskip-5mu l}}}

\def\proof{\noindent {\bf Proof.}\ \ }
\def\qed{\hfill $\Box$\medskip}

\def\textsl{{}}

\def\c0inf{C_0^\infty}
\def\proof{\noindent {\bf Proof.}\ \ }

\def\i{{\rm i}}
\newcommand{\beq}{\begin{equation}}
\newcommand{\eeq}{\end{equation}}
\newcommand{\bear}[1]{\begin{array}{#1}}
\newcommand{\ear}{\end{array}}

\newcommand{\CAR}{\mathrm{CAR}}

\renewcommand{\i}{\mathrm{i}}

\renewcommand{\d}{\mathrm{d}}

\renewcommand{\atop}[2]{\genfrac{}{}{0pt}{}{#1}{#2}}

\def\qed{$\Box$\medskip}

\def\bel{\begin{lemma}}
\def\eel{\end{lemma}}
\def\bet{\begin{theoreme}}
\def\eet{\end{theoreme}}
\def\bed{\begin{definition}}
\def\eed{\end{definition}}
\def\bep{\begin{proposition}}
\def\eep{\end{proposition}}

\def\bar{\overline}

\newcommand{\Ker}{\mathop{\mathrm{ker}}}
\newcommand{\Dom}{\mathop{\mathrm{dom}}}

\def\Ent{{\rm Ent}}

\def\fh{{\mathfrak h}}

\def\fM{{\mathfrak M}}
\def\fF{{\mathfrak F}}
\def\fG{{\mathfrak G}}

\def\dim{{\rm dim}}

\def\fr{{\rm fr}}

\begin{document}
\def\today{}
\title{On the thermodynamic limit of two-time measurement\\ entropy production}
\author{T. Benoist$^{1}$, L. Bruneau$^{2}$, V. Jak\v{s}i\'c$^{3}$, A. Panati$^4$, C.-A. Pillet$^4$
\\ \\
$^1$ Institut de Math\'ematiques de Toulouse, UMR 5219\\
Universit\'e de Toulouse, CNRS, UPS\\
31062 Toulouse Cedex 9, France
\\ \\
$^2$ Department of Mathematics\\
CY Cergy Paris University, CNRS UMR 8088\\
2 avenue Adolphe Chauvin, 95302 Cergy-Pontoise, France
\\ \\
$^3$Department of Mathematics and Statistics, McGill University\\
805 Sherbrooke Street West, Montreal,  QC,  H3A 2K6, Canada
\\ \\
$^4$Universit\'e de Toulon, CNRS, CPT, UMR 7332, 83957 La Garde, France\\
Aix-Marseille Univ, CNRS, CPT, UMR 7332, Case 907, 13288 Marseille, France
}
\maketitle
\thispagestyle{empty}
\bigskip
\centerline {\bf Dedicated to the memory of Huzihiro Araki}
\bigskip

\noindent{\small{\bf Abstract.} We provide a justification, via the thermodynamic limit, of
the modular formula for entropy production in two-time measurement proposed
in~\cite{Benoist2023a}. We consider the cases of open quantum systems in which all thermal
reservoirs are either (discrete) quantum spin systems or free Fermi gases.
}

\tableofcontents

\section{Introduction}
\label{sec-intr-new}

This work is a direct continuation of~\cite{Benoist2023a}, and we assume that
the reader is familiar with the conceptual framework, notation, and results of
this reference.

Let $(\cO,\tau,\omega)$ be a modular $C^\ast$-dynamical system
satisfying the regularity assumptions~{\bf (Reg1)} and~{\bf (Reg2)}
of~\cite{Benoist2023a}. Then, by Theorem~1.3 in this reference, for all
$\nu\in\cN$, $t\in\rr$ and $\alpha\in\i\rr$, the limit
\beq
\fF_{\nu,t}(\alpha):=\lim_{R\to\infty}\frac{1}{R}\int_0^R
\nu\left(\varsigma_\omega^\theta\left([D\omega_{-t}: D\omega]_{\alpha}\right)\right)\d\theta
\label{emm-new}
\eeq
exists, and there exists unique Borel probability measure $Q_{\nu,t}$ on $\rr$ such that
\beq
\fF_{\nu,t}(\alpha)=\int_\rr\e^{-\alpha s}\d Q_{\nu,t}(s).
\label{han-ajde-new}
\eeq
The family $(Q_{\nu,t})_{t\in\rr}$ describes the statistics of the two-time
measurement entropy production\footnote{Abbreviated 2TMEP in the sequel.} of
$(\cO,\tau,\omega)$ with respect to $\nu$. We recall that $\cN$
denotes the set of all  $\omega$-normal states on $\cO$, $\varsigma_\omega$ is
the modular group of $\omega$, $\omega_t=\omega\circ\tau^t$, and
$[D\omega_{-t}:D\omega]_\alpha$ is the Connes cocycle of the pair of states
$(\omega_{-t},\omega)$.

In the special case of a finite quantum system, $Q_{\nu,t}$ is indeed the law
of the entropy production of the system as defined by the two-time measurement
protocol of the entropic observable $-\log\omega$, assuming that at the instant
of the first measurement the system was in the state $\nu$; see Section~1.3
in~\cite{Benoist2023a}. The formulas~\eqref{emm-new}--\eqref{han-ajde-new} arise through
the customary route of modular generalization. As repeatedly emphasized
in~\cite{Benoist2023a}, this generalization requires the underlying thermodynamic
limit\footnote{In the sequel abbreviated TDL.} to be  justified on solid
physical grounds. In this work, we carry out this justification for two
paradigmatic classes of open quantum systems describing a finite quantum system
coupled to several independent thermal reservoirs.

Our starting point is an abstract TDL scheme described in Section~\ref{sec-abstract}.
This scheme is motivated by the specific  models we will
consider and has its roots in Araki's results on the continuity of the modular
structure obtained in~\cite[Section 2]{Araki1974b}, and in the specific form of the
Araki--Wyss GNS-representation of CAR-algebras induced by a quasi-free state~\cite{Araki1964a}
\footnote{See~\cite{Derezinski2013,Jaksic2010b} for pedagogical introductions to this
topic.}. One novel aspect of this scheme is the spectral assumption on the
modular operators that, in our specific settings, forces the
dynamical ergodicity assumption on thermal reservoirs. We remark that the same
ergodicity assumption is required in the main stability result of~\cite{Benoist2023a}.

The two specific settings to which we will apply our abstract TDL scheme are
{\sl Open Quantum Spin Systems,} abbreviated OQ2S, and {\sl Electronic Black Box
Models,} abbreviated EBBM. We comment on them separately.

OQ2S were introduced in~\cite{Ruelle2001}, in the study of entropy production in
non-equilibrium quantum statistical mechanics. It is in this setting that we
will make use of Araki's continuity results~\cite{Araki1974b}. As emphasized in~\cite{Ruelle2001},
the fully interacting nature of the lattice spin thermal reservoirs
brings to the forefront the role of the so-called  {\sl "boundary terms"} in the
characterization of the KMS-states by the Araki--Gibbs Condition. Due to our current
lack of understanding of the effect of these boundary terms, our results in
the OQ2S setting are incomplete and a number of important questions touching
on foundations of quantum statistical mechanics remain open. We will comment
on some of them in Section~\ref{sec-discussion-spin}.

EBBM describe open quantum systems consisting of an
electronic gas in the tight binding approximation, interacting only in a finite
subset of its countably infinite set of particle sites. Each thermal reservoir
is a free Fermi gas and the
local nature  of the  interaction makes the model amenable to rigorous analysis.
Boundary terms play no role in free Fermi gas reservoirs and our results for the
EBBM are complete. The above mentioned Araki--Wyss representation of a free Fermi
gas plays a central role in our analysis and allows for a relatively effortless
verification of the assumptions of our abstract TDL scheme. The literature on
the EBMM and related models is vast, and an incomplete list of mathematically
rigorous works on the subject is~\cite{Davies1974, Spohn1978b, Botvich1983,
Jaksic2002b, Jaksic2002a,  Froehlich2003a, Froehlich2003, Aschbacher2006,
Jaksic2006f, Jaksic2006a, Jaksic2007a, Jaksic2007b, Jaksic2010b, Cornean2014}.

The paper is organized as follows. Our abstract TDL scheme is described in
Section~\ref{sec-abstract}. In Section~\ref{sec-main-oqss} this  scheme is
applied to OQ2S and in Section~\ref{sec-main-ebbb} to EBBM. The proofs
follow the statements of the results. Sections~\ref{sec-discussion-spin}
and~\ref{sec-remarks-ebbm}, where we comment on the obtained results,
are an important part of this work.

\paragraph*{Acknowledgments} The work of CAP and VJ  was partly  funded by  the
CY Initiative grant "Investissements d'Avenir", grant number ANR-16-IDEX-0008.
The work of TB was funded by the ANR project  “ESQuisses”, grant number
ANR-20-CE47-0014-01, and  by the ANR project  “Quantum Trajectories”, grant
number ANR-20-CE40-0024-01. VJ acknowledges the support of NSERC. A part of this
work was done during long term visits of LB and AP  to McGill and CRM-CNRS
International Research Laboratory IRL 3457 at University  of Montreal. The LB
visit was funded by the CNRS and AP visits by the CRM Simons and
FRQNT-CRM-CNRS programs.

\section{An abstract TDL scheme}
\label{sec-abstract}
\subsection{A general scheme}

We denote by $(\cH,\pi,\Omega)$ the GNS representation of the $C^\ast$-algebra $\cO$ induced by the
modular state $\omega$ and write $A$ for $\pi(A)$ whenever the meaning is clear
within the context. $\fM=\pi(\cO)^{\prime\prime}$ is the enveloping von~Neumann
algebra and we denote by $\cH^+$ and $J$ the natural cone and the modular
conjugation of the pair $(\fM,\Omega)$. The set $\cN$ of $\omega$-normal
states is identified with the set of density matrices on $\cH$ and the state
$\omega$ with the vector state $(\Omega,\,\cdot\;\Omega)$. $\Delta_\nu$
denotes the modular operator of $\nu\in\cN$, $\Delta_{\nu|\mu}$ the
relative modular operator of a pair $(\nu,\mu)$ of $\omega$-normal states.
Whenever both $\nu$ and $\mu$ are faithful on $\fM$, the associated
Connes cocycle is given by
$[D\nu:D\mu]_{\i s}=\Delta_{\nu|\mu}^{\i s}\Delta_\mu^{-\i s}\in\fM$.

Let $\cNo$ be the set of states of the form $\nu_B(\cdot)=\langle B\Omega,
\,\cdot\; B\Omega\rangle$ where $B\in\fM^\prime$ and $\|B\Omega\|=1$. Since
$\Omega$ is a cyclic vector for the von~Neumann algebra $\fM^\prime$, $\cNo$ is
norm-dense in $\cN$. Our abstract approximation scheme concerns the
justification of the formula~\eqref{emm-new}--\eqref{han-ajde-new} for
$\nu=\nu_B\in\cNo$ by a themodynamic limit. Under a mild regularity assumption,
the time-evolved reference states $\omega_s=\omega \circ\tau^s$ are in $\cNo$.

\bep\label{sl-exam}
Suppose that for some $s\in\rr$ the map
$$
\i\rr\ni z\mapsto[D\omega_s:D\omega]_z\in\fM
$$
has an analytic continuation to the vertical strip $0<\Re z<\frac12$ that
is continuous and bounded  on its closure. Then $\omega_s=\nu_{B_s}\in\cNo$, with
\[
B_s=J [D\omega_s:D\omega]_{\frac12}J\in\fM'.
\]
\eep
\proof Let $\Omega_s$ be the vector representative of $\omega_s$ in the natural
cone $\cH^+$. Then
\[
\Delta_{\omega_s|\omega}^{\frac12}\Omega=\Omega_s,
\]
and so the function
\[
\i\rr\ni z\mapsto\Delta_{\omega_s|\omega}^z\Omega\in\cH
\]
has an analytic continuation to the strip $0<\Re z<\frac12$ that is continuous
and bounded on its closure; see \cite[Lemma 3]{Araki1973}.  Note that for $z\in\i\rr$,
\beq
[D\omega_s:D \omega]_z\Omega=\Delta_{\omega_s|\omega}^z\Omega,
\label{htt-1}
\eeq
and so by the Privalov theorem, see~\cite[Section III.D]{Koosis1998}, \eqref{htt-1} holds
for $0\leq\Re z\leq\frac12$. In particular,
\[
[D\omega_s:D\omega]_{\frac12}\Omega=\Delta_{\omega_s|\omega}^{\frac12}\Omega=\Omega_s.
\]
Since $J\Omega=\Omega$ and $J\Omega_s=\Omega_s$,
\[
J[D\omega_s:D \omega]_{\frac12}J\Omega=\Omega_s,
\]
and  the statement follows.\hfill\qed

Proposition~\ref{sl-exam} motivates our first assumption.

\begin{quote}{\bf (TDL1)}
For all $s\in\rr$ the map
\[
\i\rr\ni z\mapsto[D\omega_s:D\omega]_z\in\fM
\]
has an analytic continuation to the vertical strip $0<\Re z<\frac12$ that is
continuous and bounded  on its closure.
\label{tdl1}
\end{quote}
\newcommand{\TDLOne}{{\hyperref[tdl1]{{\rm (TDL1)}}}}

We consider a net of {\sl finite-dimensional} quantum dynamical systems
$(\cO_\Lambda,\tau_\Lambda,\omega_\Lambda)_{\Lambda\in\cI}$, where $\cI$ is a
directed set endowed with the order relation $\subseteq$. We denote by
$(\cH_\Lambda,\pi_\Lambda,\Omega_\Lambda)_{\Lambda\in\cI}$ the associated net of
GNS-representations. We further write
$\omega_{\Lambda,t}=\omega_\Lambda\circ\tau_\Lambda^t$ and make the following
assumptions:

\begin{quote}{\bf (TDL2)}
\ben
\item For all $\Lambda\in\cI$, $\cO_\Lambda\subseteq\cO$ and\footnote{We write $\omega>0$ to denote faithfulness of a state $\omega$. In case $\omega$ is described by a density matrix, this is equivalent to the injective character of the latter.} $\omega_\Lambda>0$.
\item If $\Lambda\subseteq\Lambda^\prime$, then $\cO_\Lambda\subseteq\cO_{\Lambda^\prime}$.
Moreover,
\[
\cO_\mathrm{loc}:=\bigcup_{\Lambda\in\cI}\cO_\Lambda
\]
is dense in $\cO$.
\item For all $A\in\cO_\mathrm{loc}$, $\lim_\Lambda\omega_\Lambda(A)=\omega(A)$.
\item For all $A\in\cO_\mathrm{loc}$, $\slim_\Lambda\tau_\Lambda^t(A)=\tau^t(A)$, locally uniformly
for $t\in\rr$.\footnote{Here and in the following $\slim_\Lambda$
denotes a strong limit, {\sl i.e.,} $\lim_\Lambda\tau_\Lambda^t(A)\Psi=\tau^t(A)\Psi$ for all $\Psi\in\cH$.}
\item For all $\Lambda\in\cI$, $\cH_\Lambda\subseteq\cH$, and $\Omega_\Lambda=\Omega$.
\item For all $t\in\rr$ and $\alpha\in\i\rr$,
\[
\slim_\Lambda\,[D\omega_{\Lambda,t}:D\omega_\Lambda]_\alpha=[D\omega_{t}:D\omega]_\alpha.
\]
\een
\label{tdl2}
\end{quote}
\newcommand{\TDLTwo}{{\hyperref[tdl2]{{\rm (TDL2)}}}}

We denote by $\Delta_{\omega_\Lambda}$  and $J_\Lambda$ the modular operator
and the modular conjugation of the pair $(\pi_\Lambda(\cO_\Lambda)'', \Omega)$,
and extend them to $\cH$ by setting $\Delta_{\omega_\Lambda}$ to be the identity
on $\cH_\Lambda^\perp$ and $J_\Lambda$ an arbitrary anti-unitary involution on
$\cH_\Lambda^\perp$. Our next assumption is

\begin{quote}{\bf (TDL3)}
\ben
\item $\slim_\Lambda\Delta_{\omega_\Lambda}^{\i\theta}=\Delta_\omega^{\i\theta}$ for all $\theta\in\rr$, and $\slim_\Lambda  J_\Lambda=J$.
\item For all $\Lambda\in\cI$, $\Ker\log\Delta_\omega\subseteq\Ker\log\Delta_{\omega_\Lambda}$.
\een
\label{tdl3}
\end{quote}
\newcommand{\TDLThree}{{\hyperref[tdl3]{{\rm (TDL3)}}}}

This assumption requires a comment. As we shall see, \TDLThree(2) holds
automatically for open quantum systems in which each thermal reservoir is an
ergodic quantum dynamical system. Regarding~\TDLThree(1), the following set
of results is established in~\cite[Section 2]{Araki1974b}.\footnote{For an erratum,
see~\cite[Section 5, Remark 2]{Araki1975/76}.}

\begin{theorem}\label{araki74}
Suppose that~\TDLTwo{\rm (1--2)} hold, and that for all $\Lambda\in\cI$,
$$
\omega_\Lambda=\omega\big|_{\cO_\Lambda},\qquad
\pi_\Lambda=\pi\big|_{\cO_\Lambda},\qquad
\Omega_\Lambda=\Omega,
$$
where $(\,\cdot\,)\big|_{\cO_\Lambda}$ denotes the restriction from $\cO$ to $\cO_\Lambda$.
Then the following hold:
\ben
\item $\omega(A)=\omega_\Lambda(A)$ for $A\in\cO_\Lambda$.
\item $\slim_\Lambda\Delta_{\omega_\Lambda}^{\i\theta}=\Delta_{\omega}^{\i\theta}$, locally uniformly for $\theta\in\rr$.
\item $\slim_\Lambda J_\Lambda= J$.
\item Let $(Q_\Lambda)_{\Lambda\in\cI}$ with $Q_\Lambda\in\cO_\Lambda$ be such that,
for some $Q\in\cO$,
$$
\slim_\Lambda  Q_\Lambda=Q,\qquad\slim_\Lambda  Q_\Lambda^\ast =Q^\ast.
$$
Then,
\[
\lim_\Lambda\Delta_{\omega_\Lambda}^z Q_\Lambda\Omega=\Delta_\omega^z Q\Omega,
\]
locally uniformly for $z$ in the vertical strip $0\leq\Re z\le\frac12$.
\een
\end{theorem}

We note in particular that~\TDLTwo(3) and~\TDLThree(1) hold under the assumptions of Theorem~\ref{araki74}.

The final assumption in this section is:
\newcommand{\TDLFour}{{\hyperref[tdl4]{{\rm (TDL4)}}}}
\begin{quote}{\bf (TDL4)}
\TDLOne{} holds and for all $s\in \rr$,
\begin{align*}
\slim_\Lambda\,[D\omega_{\Lambda, s}: D\omega_\Lambda]_\frac{1}{2}
&=[D\omega_{s}: D\omega]_\frac{1}{2},\\[4pt]
\slim_\Lambda\,[D\omega_{\Lambda, s}: D\omega_\Lambda]_\frac{1}{2}^\ast
&=[D\omega_{s}: D\omega]_\frac{1}{2}^\ast.
\end{align*}
\label{tdl4}
\end{quote}
For each $\nu_B\in\cNo$ we set $\nu_{B,\Lambda}:=\nu_B\big|_{\cO_\Lambda}$. The 2TMEP of
$(\cO_\Lambda,\tau_\Lambda,\omega_\Lambda)$ with respect to $\nu_{B,\Lambda}$ is
defined by the formulas~\eqref{emm-new}--\eqref{han-ajde-new},
$$
\fF_{\nu_{B,\Lambda},t}(\alpha)
=\lim_{R\to\infty}\frac1R\int_0^R\nu_{B,\Lambda}\left(
\varsigma_{\omega_\Lambda}^\theta\left([D\omega_{\Lambda, -t}:D\omega_\Lambda]_{\alpha}\right)
\right)\d\theta,
$$
$$
\fF_{\nu_{B,\Lambda},t}(\alpha)=\int_\rr\e^{-\alpha s}\d Q_{\nu_{B,\Lambda},t}(s).
$$
Of course, in this case $Q_{\nu_{B,\Lambda},t}$ is just the law of the 2TMEP of $\omega_\Lambda$,
the system being in the state $\nu_{B,\Lambda}$ at the instant of the first
measurement; see Section~1.3 in~\cite{Benoist2023a}. If $\nu_{B, \Lambda}$ is replaced
with $\omega_{\Lambda,s}=\omega_\Lambda\circ\tau_\Lambda^s$, then
$\fF_{\omega_{\Lambda,s},t}$ and $Q_{\omega_{\Lambda,s},t}$
describe the 2TMEP of $\omega_\Lambda$, the system being in the state
$\omega_{\Lambda,s}$ at the instant of the first measurement. This second case
is of  importance for our study of a quantum Gallavotti--Cohen fluctuation
theorem in~\cite{Benoist2024}. Note that, except in trivial cases,
$\omega_s\big|_{\cO_\Lambda}\neq\omega_{\Lambda, s}$.

The main result in this section is :

\begin{theorem} Suppose that~\TDLTwo~and~\TDLThree{} hold. Then,
\ben
\item Given any state $\nu_B\in\cN_0$, for all $t\in\rr$ and $\alpha\in\i\rr$ one has
\beq
\lim_\Lambda\fF_{\nu_{B,\Lambda},t}(\alpha)=\fF_{\nu_B,t}(\alpha),
\label{pasulj-1}
\eeq
and $\lim_\Lambda Q_{\nu_{B,\Lambda},t}=Q_{\nu_B,t}$ weakly.

\item  Suppose in addition that~\TDLFour{} holds. Then
for all $s,t\in\rr$ and $\alpha\in\i\rr$ one has
\beq
\lim_\Lambda\fF_{\omega_{\Lambda,s},t}(\alpha)=\fF_{\omega_s,t}(\alpha),
\label{pasulj-2}
\eeq
and  $\lim_\Lambda Q_{\omega_{\Lambda,s},t}=Q_{\omega_s,t}$ weakly.
\een
\label{ss-sunday}
\end{theorem}

\proof {\bf (1)} Central to the argument are the  formulas
\beq
\begin{split}
\fF_{\nu_B,t}(\alpha)
&=\langle B^\ast B\Omega,P[D\omega_{-t}:D\omega]_\alpha\Omega\rangle,\\[4pt]
\fF_{\nu_{B,\Lambda},t}(\alpha)
&=\langle B^\ast B\Omega,P_\Lambda[D\omega_{\Lambda,-t}:D\omega_\Lambda]_\alpha\Omega\rangle,
\end{split}
\label{ssd}
\eeq
where $P$ and $P_\Lambda$ denote the orthogonal projections onto $\ker\log\Delta_\omega$ and
$\ker\log\Delta_{\omega_\Lambda}$ respectively. The formulas~\eqref{ssd} hold for all $\nu_B\in\cNo$ and
are  easy consequences of the fact that $B\in\fM'$; see the proof of Theorem~1.3
in~\cite{Benoist2023a}. By~\TDLTwo(6) and the polarization identity,
to prove~\eqref{pasulj-1} it suffices to show that for all $\Psi\in\cH$,
\beq
\lim_\Lambda\langle\Psi,P_\Lambda\Psi\rangle=\langle\Psi,P\Psi\rangle.
\label{s-new}
\eeq
Assumption~\TDLThree(2) gives that $\langle\Psi,P_\Lambda\Psi\rangle\geq\langle \Psi,P\Psi\rangle$
for all $\Lambda\in\cI$, and so
\beq
\liminf_\Lambda\langle\Psi,P_\Lambda\Psi\rangle\ge\langle\Psi,P\Psi\rangle.
\label{ss-new}
\eeq
Theorem~\ref{araki74}(2) gives that, for $t\in\rr$,
$$
\lim_\Lambda\langle\Psi,\Delta_{\omega_\Lambda}^{\i t}\Psi\rangle=\langle\Psi,\Delta_{\omega}^{\i t}\Psi\rangle.
$$
By  L\'evy's continuity theorem, the spectral measure $\mu_{\Psi,\Lambda}$ of
$\log\Delta_{\omega_\Lambda}$ for the vector $\Psi$ converges weakly to the spectral
measure $\mu_{\Psi}$ of $\log \Delta_{\omega}$ for the same vector $\Psi$.
By the Portmanteau Theorem~\cite[Theorem~8.2.3]{Bogachev2007}, this convergence gives
\[
\limsup_\Lambda\mu_{\Psi,\Lambda}(\{0\})\leq\mu_{\Psi}(\{0\}),
\]
and so
\beq
\limsup_\Lambda\langle\Psi,P_\Lambda\Psi\rangle\le\langle \Psi,P\Psi\rangle.
\label{sss-new}
\eeq
The inequalities~\eqref{ss-new} and~\eqref{sss-new} yield~\eqref{s-new}. The second
claim follows from~\eqref{pasulj-1} and L\'evy's continuity theorem.

\medskip\noindent{\bf (2)} Writing $C_\Lambda=[D\omega_{\Lambda,s}:D\omega_\Lambda]_{\frac12}$ and
$C=[D\omega_s: D\omega]_{\frac12}$, it follows from~\eqref{ssd} and
Proposition~\ref{sl-exam} that
\begin{align*}
\fF_{\omega_{\Lambda,s},t}(\alpha)
&=\langle J_\Lambda C_\Lambda^\ast C_\Lambda\Omega,P_\Lambda[D\omega_{\Lambda,-t}:D\omega_\Lambda]_\alpha\Omega\rangle,\\[4pt]
\fF_{\omega_s,t}(\alpha)
&=\langle JC^\ast C\Omega,P[D\omega_{-t}:D\omega]_\alpha\Omega\rangle.
\end{align*}
By~\TDLThree(1), the isometric modular conjugations are strongly convergent, and
by~\TDLFour{} and the uniform boundedness principle, one has
$\sup_\Lambda\|C_\Lambda\|<\infty$. Thus, it follows from the telescopic expansion
$$
J_\Lambda C_\Lambda^\ast C_\Lambda\Omega-JC^\ast C\Omega=(J_\Lambda-J)C^\ast C\Omega
+J_\Lambda(C_\Lambda^\ast-C^\ast)C\Omega
+J_\Lambda C_\Lambda^\ast(C_\Lambda-C)\Omega,
$$
that
\[
\lim_\Lambda J_\Lambda C_\Lambda^\ast C_\Lambda\Omega=JC^\ast C\Omega,
\]
while by~\TDLTwo(6),
\[
\lim_\Lambda\,[D\omega_{\Lambda,-t}:D\omega_\Lambda]_\alpha\Omega
=[D\omega_{-t}:D\omega]_\alpha\Omega.
\]
Invoking~\eqref{s-new} again gives~\eqref{pasulj-2} and L\'evy's continuity theorem
yields the second claim.\hfill\qed

\subsection{Thermodynamic limit of Connes' cocycles}
\label{sec-cocycles}

In this section we introduce more structure to the dynamical system
$(\cO,\tau,\omega)$. The purpose of these new elements is to allow us to exploit
the structural stability of KMS states, as embodied in Araki's perturbation
theory, see~\cite[Section~5.4]{Bratteli1981}. Quite unexpectedly, this part of
the algebraic theory of equilibrium quantum statistical mechanics developed in
the 70', is also at the hearth of some more recent advances in nonequilibrium
quantum statistical mechanics. The next two sets of assumptions will allow us to
use the concept of entropy production in nonequilibrium processes to gain
control on the Connes cocycles. This will lead us to check the corresponding
Assumptions~\TDLOne, \TDLTwo(6) and~\TDLFour.

We denote by $\delta_\omega$ the $\ast$-derivation generating the modular group
$\varsigma_\omega$: $\varsigma^\theta_\omega=\e^{ \theta\delta_\omega}$.
Our next assumption is:
\newcommand{\TDLFive}{{\hyperref[tdl5]{{\rm (TDL5)}}}}
\begin{quote}{\bf (TDL5)}
\ben
\item $\varsigma_\omega$ commutes with a ``free'' $C^\ast$-dynamics $\tau_\fr^t=\e^{t\delta_\fr}$
which leaves the state $\omega$ invariant:
$$
\varsigma_\omega^\theta\circ\tau_\fr^t=\tau_\fr^t\circ\varsigma_\omega^\theta,\qquad
\omega\circ\tau_\fr^t=\omega,
$$
for all $\theta,t\in\rr$.
\item $\tau$ is a {\sl local perturbation} of $\tau_\fr$:
$$
\tau^t=\e^{t\delta},\qquad
\delta=\delta_\fr+\i[V,\,\cdot\;],
$$
where $V$ is a self-adjoint element of $\cO$.
\item $V\in\Dom(\delta_\omega)$, and we define
$$
\sigma=\delta_\omega(V).
$$
\item The map
\[
\rr\ni\theta\mapsto\varsigma_{\omega}^\theta(\sigma)\in\cO
\]
has an analytic continuation to the horizontal strip $|\Im\theta|<\frac{1}{2}$ that
is bounded and continuous on its closure.
\een
\label{tdl5}
\end{quote}
Starting with the works~\cite{Jaksic2001a, Ruelle2001}, the operator $\sigma$ has played
an important role in many developments in non-equilibrium quantum statistical mechanics as
the {\sl entropy production observable} of the $C^\ast$-dynamical system $(\cO,\tau,\omega)$.
In the sequel, we denote by $\cL$ (respectively $\cL_\fr$) the standard Liouvillean of the dynamics
$\tau$ (respectively $\tau_\fr$), {\sl i.e.,} the unique self-adjoint operator on $\cH$ such that
$$
\pi(\tau^t(\,\cdot\,))=\e^{\i t\cL}\pi(\,\cdot\,)\e^{-\i t\cL},\qquad\e^{-\i t\cL}\cH^+\subset\cH^+,
$$
for all $t\in\rr$ (and similarly for the free dynamics).

\begin{lemma}\label{lem:relDeltaForm}
Under the assumptions~\TDLFive{\rm (1--3)}, one has
\[
\log\Delta_{\omega_s|\omega}=\log\Delta_\omega+Q_s,\qquad Q_s=\int_0^s\tau^{-t}(\sigma)\d t,
\]
for any $s\in\rr$.
\end{lemma}

\proof A proof of the statement is implicit in~\cite{Jaksic2003}. For the reader's
convenience and later references, we sketch the argument. Let $(\Gamma_s)_{s\in\rr}$
be the cocycle associated to the local perturbation $V$ of the free dynamics $\tau_\fr$,
{\sl i.e.,} the solution of the Cauchy problem
\beq
\partial_s\Gamma_s=\i\Gamma_s\tau_\fr^s(V),\qquad\Gamma_0=\one.
\label{equ:GammaDef}
\eeq
$\Gamma_s$ is a unitary element of $\cO$ with the norm convergent expansion
\beq
\Gamma_s=\one+\sum_{n\ge1}\i^n\int\limits_{0\le s_1\le\cdots\le s_n\le s}
\tau_\fr^{s_1}(V)\cdots\tau_\fr^{s_n}(V)\d s_1\cdots\d s_n,
\label{equ:GammaExp}
\eeq
and such that, as a consequence of~\TDLFive(2),
\beq
\tau^s(\,\cdot\,)=\Gamma_s\tau_\fr^s(\,\cdot\,)\Gamma_s^\ast,
\label{equ:tauForm}
\eeq
see, e.g., \cite[Section~5.4.1]{Bratteli1981}. From the perspective of the associated
$W^\ast$-dynamics, one has
$$
\cL=\cL_\fr+\pi(V)-J\pi(V)J,
$$
and
$$
\e^{\i s\cL}=J\pi(\Gamma_s)J\pi(\Gamma_s)\e^{\i s\cL_\fr}.
$$
Taking~\TDLFive(1) into account, we have $\e^{\i s\cL_\fr}\Omega=\Omega$, so that
the vector representative of the state $\omega_s$ in the natural cone $\cH^+$ is
\beq
\Omega_s=\e^{-\i s\cL}\Omega=J\pi(\Gamma_{-s})J\pi(\Gamma_{-s})\Omega.
\label{equ:OmegasForm}
\eeq

For some arbitrary but fixed $s\in\rr$ and any $\theta\in\rr$, set
\beq
T_\theta=\Gamma_{-s}\varsigma_\omega^\theta(\Gamma_{-s}^\ast).
\label{equ:TTdef}
\eeq
It follows from~\TDLFive(3) and the expansion~\eqref{equ:GammaExp} that
$\Gamma_{-s}\in\Dom(\delta_\omega)$. Using~\eqref{equ:GammaDef}, one easily checks that
$$
\partial_\theta T_\theta=\i T_\theta\varsigma_\omega^\theta(Q_{s}),\qquad T_0=\one,
$$
where $Q_s=\i\delta_\omega(\Gamma_{-s})\Gamma_{-s}^\ast=Q_s^\ast\in\cO$. Thus,
$(T_\theta)_{\theta\in\rr}$ is the unitary cocycle associated to the local
perturbation $Q_s$ of the modular group
$$
\alpha^\theta(\,\cdot\,)=\e^{\theta(\delta_\omega+\i[Q_s,\,\cdot\;])}
=T_\theta\varsigma_\omega^\theta(\,\cdot\,)T_\theta^\ast.
$$
Recalling that
\[
\pi(\varsigma_\omega^\theta(\,\cdot\,))=\e^{\i\theta\log\Delta_\omega}\pi(\, \cdot\,)\e^{-\i\theta\log \Delta_\omega},
\]
we derive
$$
\pi(\alpha^\theta(\,\cdot\,))
=\pi(T_\theta)\e^{\i\theta\log\Delta_\omega}\pi(\, \cdot\,)\e^{-\i\theta\log \Delta_\omega}\pi(T_\theta^\ast).
$$
Thus, $\alpha$ extends to a $W^\ast$-dynamics on $\fM$ which we also denote by $\alpha$,
$$
\alpha^\theta(\,\cdot\,)=\e^{\i\theta(\log\Delta_\omega+Q_{s})}(\,\cdot\,)\e^{-\i\theta(\log\Delta_\omega+Q_{s})},
\qquad \e^{\i\theta(\log\Delta_\omega+Q_{s})}=\pi(T_\theta)\e^{\i\theta\log\Delta_\omega}.
$$
Using~\eqref{equ:TTdef} we obtain, for any $A\in\cO$ and $\theta\in\rr$,
$$
\e^{\i\theta(\log\Delta_\omega+Q_{s})}\pi(A)\Omega=\pi(T_\theta)\e^{\i\theta\log\Delta_\omega}\pi(A)\Omega
=\pi(\Gamma_{-s})\e^{\i\theta\log\Delta_\omega}\pi(\Gamma_{-s}^\ast A)\Omega.
$$
Since the right-hand side has an analytic continuation to $\theta=-\i/2$, we further get
$$
\e^{(\log\Delta_\omega+Q_{s})/2}\pi(A)\Omega=\pi(\Gamma_{-s})\Delta_\omega^{\frac12}\pi(\Gamma_{-s}^\ast A)\Omega
=\pi(\Gamma_{-s})J\pi(A^\ast\Gamma_{-s})\Omega,
$$
and multiplication on the left by $J$ yields, taking~\eqref{equ:OmegasForm} into account,
$$
J\e^{(\log\Delta_\omega+Q_{s})/2}\pi(A)\Omega
=J\pi(\Gamma_{-s})J\pi(A^\ast\Gamma_{-s})\Omega
=\pi(A^\ast)J\pi(\Gamma_{-s})J\pi(\Gamma_{-s})\Omega
=\pi(A^\ast)\Omega_s,
$$
which shows that $\log\Delta_\omega+Q_{s}=\log\Delta_{\omega_s|\omega}$.
To finish the proof, we note that a simple calculation using~\eqref{equ:GammaDef} yields
$$
\partial_sQ_s=\tau^{-s}(\sigma),
$$
so that
$$
Q_{s}=\int_0^s\tau^{-t}(\sigma)\d t.
$$
\hfill\qed

\bep\label{sd-new}
Assumption~\TDLFive{} implies~\TDLOne.
\eep

\proof We shall use the notation and intermediate results from the previous proof.
By Lemma~\ref{lem:relDeltaForm}, the cocycle
\[
[D\omega_s: D\omega]_z=\e^{z \log \Delta_{\omega_s|\omega}}\e^{-z\log \Delta_\omega}
\]
has, for $z\in\i\rr$,  the norm convergent expansion
\beq
[D\omega_s: D\omega]_z=\one+ \sum_{n\geq 1} z^n \int\limits_{0\leq \theta_1\leq \cdots\leq \theta_n\leq 1}
\varsigma_{\omega}^{-\i \theta_1 z}(Q_s)\cdots
\varsigma_\omega^{-\i \theta_n z}(Q_s)\d \theta_1\cdots \d \theta_n.
\label{st-new}
\eeq
Using~\eqref{equ:tauForm}, we have for $\theta\in\rr$,
$$
\varsigma_\omega^{\theta}(Q_s)=\int_0^s\varsigma_\omega^{\theta}(\tau^{-t}(\sigma))\d t
=\int_0^s\varsigma_\omega^{\theta}(\Gamma_{-t}\tau_\fr^{-t}(\sigma)\Gamma_{-t}^\ast)\d t,
$$
and since the modular group commutes with the free dynamics, we can write
\beq
\varsigma_\omega^{\theta}(Q_s)=\int_0^s\varsigma_\omega^{\theta}(\Gamma_{-t})
\tau_\fr^{-t}(\varsigma_\omega^{\theta}(\sigma))\varsigma_\omega^{\theta}(\Gamma_{-t}^\ast)\d t.
\label{equ:sigQs}
\eeq
By Assumption~\TDLFive(4), $\theta\mapsto\varsigma_\omega^{\theta}(\sigma)$ is analytic in the strip
$|\Im\theta|<\frac12$, and bounded and continuous on its closure. Since
$\partial_\theta\varsigma_\omega^\theta(V)=\varsigma_\omega^\theta(\sigma)$, the same is true for
$$
\theta\mapsto\varsigma_{\omega}^\theta(V)=V+\theta\int_0^1\varsigma_\omega^{\theta t}(\sigma)\d t.
$$
That the same is also true for  $\theta\mapsto\varsigma_{\omega}^\theta(\Gamma_{-t})$ is a
consequence of the expansion~\eqref{equ:GammaExp}, since the latter implies
\beq
\varsigma_\omega^{\theta}(\Gamma_{-t})=\one+\sum_{n\ge1}(-\i t)^n\int\limits_{0\le s_1\le\cdots\le s_n\le1}
\tau_\fr^{-ts_1}(\varsigma_{\omega}^\theta(V))\cdots\tau_\fr^{-ts_n}(\varsigma_{\omega}^\theta(V))\d s_1\cdots\d s_n.
\label{equ:sigomegagamma}
\eeq
Invoking~\eqref{equ:sigQs}, we conclude that the function $\theta\mapsto\varsigma_{\omega}^\theta(Q_s)$
is analytic in the strip $|\Im\theta|<\frac12$, and bounded and continuous on its closure. Finally, from the
expansion~\eqref{st-new} we conclude that the map
$$
z\mapsto[D\omega_s:D\omega]_z
$$
is analytic on the strip $|\Re z|<\frac12$, and bounded and continuous on its closure.\hfill \qed

In a similar spirit, we will assume the following additional properties of the approximation scheme
we have introduced in Section~\ref{sec-abstract}.

\newcommand{\TDLSix}{{\hyperref[tdl6]{{\rm (TDL6)}}}}
\begin{quote}{\bf (TDL6)}~\TDLTwo(5) and \TDLFive{} hold. Moreover, for all $\Lambda\in\cI$:
\ben
\item $\varsigma_{\omega_\Lambda}$ commutes with a ``free'' $C^\ast$-dynamics
$\tau_{\Lambda,\fr}^t=\e^{t\delta_{\Lambda,\fr}}$ which leaves the state $\omega_\Lambda$ invariant:
$$
\varsigma_{\omega_\Lambda}^\theta\circ\tau_{\Lambda,\fr}^t
=\tau_{\Lambda,\fr}^t\circ\varsigma_{\omega_\Lambda}^\theta,\qquad \omega_\Lambda\circ\tau_{\Lambda,\fr}^t=\omega_\Lambda,
$$
for all $\theta, t\in\rr$.
\item $\tau_\Lambda$ is a {\sl local perturbation} of $\tau_{\Lambda,\fr}$:
$$
\tau_\Lambda^t=\e^{t\delta_\Lambda},\qquad
\delta_\Lambda=\delta_{\Lambda,\fr}+\i[V_\Lambda,\,\cdot\;],
$$
where $V_\Lambda$ is a self-adjoint element of $\cO_\Lambda$ such that
$$
\lim_\Lambda V_\Lambda=V
$$
holds in $\cO$.
\item Extending the standard Liouvillean $\cL_{\Lambda,\fr}$ of the free dynamics
$\tau_{\Lambda,\fr}$ to $\cH$ by setting it to $0$ on $\cH_\Lambda^\perp$, one has
$$
\slim_\Lambda\e^{\i t\cL_{\Lambda,\fr}}=\e^{\i t\cL_\fr},
$$
locally uniformly for $t\in\rr$.
\item The entropy production observable of the $C^\ast$-dynamical
system $(\cO_\Lambda,\tau_\Lambda,\omega_\Lambda)$,
$$
\sigma_\Lambda=\delta_{\omega_\Lambda}(V_\Lambda),
$$
satisfies
\[
\slim_\Lambda\varsigma_{\omega_\Lambda}^\theta(\sigma_\Lambda)=\varsigma_\omega^\theta(\sigma)
\]
locally uniformly for $|\Im\theta|\leq\frac12$.
\een
\label{tdl6}
\end{quote}

\bep\label{sd-new-new}
Assumption \TDLSix{} implies \TDLTwo{\rm (6)} and \TDLFour.
\eep
\proof Since \TDLSix(1--2) imply that the $C^\ast$-dynamical system
$(\cO_\Lambda,\tau_\Lambda,\omega_\Lambda)$ satisfies the assumptions of
Lemma~\ref{lem:relDeltaForm}, we can start, as in the proof of
Proposition~\ref{sd-new}, with the expansions
\beq
\begin{split}
[D\omega_{\Lambda,s}:D\omega_\Lambda]_z&=\one+\sum_{n\geq 1} z^n\int\limits_{0\leq \theta_1\leq\cdots\leq\theta_n\leq 1}
\varsigma_{\omega_\Lambda}^{-\i\theta_1 z}(Q_{\Lambda,s})\cdots
\varsigma_{\omega_\Lambda}^{-\i\theta_n z}(Q_{\Lambda,s})\d \theta_1\cdots \d \theta_n,\\[4pt]
[D\omega_{\Lambda,s}: D\omega_\Lambda]_z^\ast&=\one+\sum_{n\geq 1} \bar{z}^n\int\limits_{0\leq \theta_1\leq\cdots\leq\theta_n\leq 1}
\varsigma_{\omega_\Lambda}^{\i\theta_n\bar z}(Q_{\Lambda,s})\cdots
\varsigma_{\omega_\Lambda}^{\i\theta_1\bar z}(Q_{\Lambda,s})\d \theta_1\cdots \d \theta_n,
\end{split}
\label{st-new-1}
\eeq
where $Q_{\Lambda,s}=\int_0^s\tau_\Lambda^{-t}(\sigma_{\Lambda})\d t$. $\cO_\Lambda$ being
finite dimensional, the relations~\eqref{st-new-1} hold for all $z\in\cc$. We again write
\beq
\varsigma_{\omega_\Lambda}^\theta(Q_{\Lambda,s})
=\int_0^s\varsigma_{\omega_\Lambda}^{\theta}(\Gamma_{\Lambda,-t})
\tau_{\Lambda,\fr}^{-t}(\varsigma_{\omega_\Lambda}^{\theta}(\sigma_\Lambda))
\varsigma_{\omega_\Lambda}^{\theta}(\Gamma_{\Lambda,-t}^\ast)\d t,
\label{sst-new-1}
\eeq
where
\beq
\varsigma_{\omega_\Lambda}^{\theta}(\Gamma_{\Lambda,-t})
=\one+\sum_{n\ge1}(-\i t)^n\int\limits_{0\le s_1\le\cdots\le s_n\le1}
\tau_{\Lambda,\fr}^{-ts_1}(\varsigma_{\omega_\Lambda}^\theta(V_\Lambda))\cdots
\tau_{\Lambda,\fr}^{-ts_n}(\varsigma_{\omega_\Lambda}^\theta(V_\Lambda))\d s_1\cdots\d s_n.
\label{equ:sigomegalambdagamma}
\eeq
Since
$$
\varsigma_\omega^\theta(V)-\varsigma_{\omega_\Lambda}^\theta(V_\Lambda)
=V-V_\Lambda+\theta\int_0^1(\varsigma_\omega^{s\theta}(\sigma)-\varsigma_{\omega_\Lambda}^{s\theta}(\sigma_\Lambda))\d s,
$$
it follows from~\TDLSix(2)+(4) that
$$
\slim_\Lambda\varsigma_{\omega_\Lambda}^\theta(V_\Lambda)=\varsigma_\omega^\theta(V),
$$
locally uniformly for $|\Im\theta|\le\frac12$. The group property
and the isometric nature of the modular dynamics, together with the uniform boundedness principle yield
$$
\sup_{\Lambda,|\Im\theta|\le1/2}\|\varsigma_{\omega_\Lambda}^\theta(V_\Lambda)\|<\infty,\qquad
\sup_{\Lambda,|\Im\theta|\le1/2}\|\varsigma_{\omega_\Lambda}^\theta(\sigma_\Lambda)\|<\infty,
$$
and since $\e^{\i t\cL_{\Lambda,\fr}}$ is unitary, a telescopic expansion yields, as in the proof of
Theorem~\ref{ss-sunday}(2),
$$
\slim_\Lambda\varsigma_{\omega_\Lambda}^{\theta}(\Gamma_{\Lambda,-t})
=\varsigma_{\omega}^{\theta}(\Gamma_{-t}),
$$
for $|\Im\theta|\le\frac12$.
By the same argument, \eqref{sst-new-1} yields
$$
\slim_\Lambda\varsigma_{\omega, \Lambda}^\theta(Q_{\Lambda,s})=\varsigma_{\omega}^\theta(Q_{s}),
$$
and in particular
$$
\sup_{\Lambda,|\Im\theta|\le1/2}\|\varsigma_{\omega_\Lambda}^\theta(Q_{\Lambda,s})\|<\infty.
$$
Finally, we deduce from the expansion~\eqref{st-new-1} that
$$
\slim_\Lambda[D\omega_{\Lambda,s}: D\omega_\Lambda]_z=[D\omega_{s}: D\omega]_z,\qquad
\slim_\Lambda[D\omega_{\Lambda,s}: D\omega_\Lambda]_z^\ast=[D\omega_{s}: D\omega]_z^\ast,
$$
hold for all $z$ in the closed strip $|\Im z|\le\frac12$.\hfill\qed

\section{Open Quantum Spin Systems}
\label{sec-main-oqss}

\subsection{Quantum spin systems}
\label{sec-qss}
We follow~\cite{Bratteli1981}; see also~\cite{Israel1979, Simon1993, Ruelle1969}.

\noindent{\bf The $C^\ast$-algebra.} Let $G$ be a countably infinite set. At this
point, no further structure on $G$ is assumed. The collection of all finite
subsets of $G$ is denoted by $\fG_\mathrm{fin}$. Let $\fh$ be the finite
dimensional Hilbert space of a single spin. To each $x\in G$ we associate a copy
$\fh_x$ of $\fh$, and to each $\Lambda\in\fG_\mathrm{fin}$ the Hilbert space
\[
\cK_\Lambda=\bigotimes_{x\in\Lambda}\fh_x.
\]
$\cO_\Lambda$ denotes the $C^\ast$-algebra of all linear operators on
$\cK_\Lambda$. Its elements  describe observables of the spins localized in
the region $\Lambda$. For  $\Lambda\subseteq \Lambda^\prime$ one naturally identifies
$\cO_\Lambda$ with a  $C^\ast$-subalgebra of $\cO_{\Lambda^\prime}$.
The $\ast$-algebra of local observables  is
\[
\cO_\mathrm{loc}=\bigcup_{\Lambda\in\fG_\mathrm{fin}}\cO_\Lambda.
\]
Finally, the $C^\ast$-algebra of the spin system over $G$ is the norm closure $\cO_G$ of
$\cO_\mathrm{loc}$. The algebra $\cO_G$ is unital, simple, and
separable.  For any $G_0\subset G$ one has a natural identification
$\cO_G=\cO_{G_0}\otimes\cO_{G_0^\mathrm{c}}$.\footnote{We will write $A$ for
$A\otimes\one$, $\one\otimes A$ whenever the meaning is clear within the
context.} Whenever $G$ is  understood, we write $\cO$ for $\cO_G$.

\noindent{\bf  Dynamics.} An {\sl interaction} is a map
\[
\Phi : {\mathfrak G}_{\rm fin}\rightarrow {\cal O}
\]
such that $\Phi(X)$ is a self-adjoint element of $\cO_X$.
The {\sl Hamiltonian} of a region $\Lambda\in\fG_\mathrm{fin}$ is the local observable defined by
\[
H_\Lambda(\Phi)=\sum_{X\subseteq\Lambda}\Phi(X).
\]
It generates a local $C^\ast$-dynamics $\tau_{\Phi,\Lambda}$ on $\cO$, where
\[
\tau_{\Phi,\Lambda}^t(A)=\e^{\i tH_\Lambda(\Phi)}A\e^{-\i tH_\Lambda(\Phi)}.
\]
To control $\tau_{\Phi,\Lambda}$ in the thermodynamic limit $\Lambda\uparrow G$,
one needs a suitable regularity assumption. We settle for
\newcommand{\SR}{{\hyperref[sr]{{\rm (SR)}}}}
\begin{quote}{\bf (SR)}
For some $\lambda >0$,
\[
\|\Phi\|_\lambda=
\sup_{x\in G}\sum_{X\ni x}\|\Phi(X)\|\e^{\lambda(|X|-1)}<\infty,
\]
where $|X|$ denotes the cardinality of the set $X$.
\label{sr}
\end{quote}
Theorem~6.2.4 in~\cite{Bratteli1981} and its proof give the following.

\begin{theorem}\label{p-oce}
Suppose that~\SR{} holds. Then:
\ben
\item For all $A\in\cO$ the limit
\[
\tau_\Phi^t(A):=\lim_\Lambda\tau_{\Phi,\Lambda}^t(A)
\]
exists in norm, locally uniformly for $t\in\rr$.
\item $\tau_\Phi=\{\tau_\Phi^t\mid t\in\rr\}$ is a $C^\ast$-dynamics on $\cO$. We  denote by
$\delta_\Phi$ its generator.
\item $\cO_\mathrm{loc}$ is a core of $\delta_\Phi$, and  for $A\in\cO_{\Lambda}$,
\[
\delta_\Phi(A)=\sum_{X\cap\Lambda\neq\emptyset}\i[\Phi(X),A].
\]
\item For $A\in\cO_\Lambda$ and $n\geq 1$,
$$
\|\delta_\Phi^n(A)\|\le\frac{2^n n!}{\lambda^n}\e^{\lambda|\Lambda|}\|\Phi\|_\lambda^n \|A\|.
$$
In particular, for  all
$A\in\cO_\mathrm{loc}$, the map
\[
\rr\ni t\mapsto\tau_\Phi^t(A)\in\cO
\]
has an analytic extension to the strip $|\Im z|<\frac{\lambda}{2\|\Phi\|_{\lambda}}$.
\een
\end{theorem}

Until the end of this section we assume that~\SR{} holds. We will make frequent use of the fact that,
whenever $A=\sum_nA_n$ is a norm convergent series with $A_n\in\cO_\mathrm{loc}$ and
$\sum_n\|\delta_\Phi(A_n)\|<\infty$, then Property~(3)
implies that $A\in\Dom(\delta_\Phi)$ with $\delta_\Phi(A)=\sum_n\delta_\Phi(A_n)$.

\noindent{\bf KMS-states.} For $\Lambda\in\fG_\mathrm{fin}$, the {\sl local Gibbs state} on $\cO_\Lambda$,
at inverse temperature $\beta>0$, is defined by the density matrix
\beq
\omega_{\beta,\Lambda}=\frac{\e^{-\beta H_\Lambda(\Phi)}}{\tr(\e^{-\beta H_\Lambda(\Phi)})}.
\label{thgivs}
\eeq
Using the identification $\cO=\cO_\Lambda\otimes\cO_{\Lambda^\mathrm{c}}$, one extends
$\omega_{\beta,\Lambda}$ (in an arbitrary way) to  a state on $\cO$. Denoting
this extension by $\bar\omega_{\beta,\Lambda}$, any weak$^\ast$-limit point of the net
$(\bar\omega_{\beta,\Lambda})_{\Lambda\in\fG_\mathrm{fin}}$ is a $(\tau_\Phi,\beta)$-KMS
state on $\cO$ \cite[Theorem~6.2.15]{Bratteli1981}. Any  $(\tau_\Phi,\beta)$-KMS states
that arises in this way is  called a {\sl thermodynamic limit} KMS state.
This construction in particular gives that the set $\cS_{\tau_\Phi,\beta}$  of all
$(\tau_\Phi,\beta)$-KMS states is non-empty.

If it happens that $\cS_{\tau_\Phi,\beta}$ is a singleton, then the net
$(\bar\omega_{\beta,\Lambda})_{\Lambda \in\fG_\mathrm{fin}}$ converges
to the unique $(\tau_\Phi,\beta)$-KMS state. This is known to be the case in
the high-temperature regime, {\sl i.e.,} for $\beta\|\Phi\|_\lambda$ small enough.
For some concrete estimates, see~\cite[Proposition 6.2.45]{Bratteli1981},
or~\cite{Froehlich2015} for more recent results.

If $\cS_{\tau_\Phi,\beta}$ is not a singleton, one needs to supply boundary conditions
to the local Gibbs states in order to reach all $(\tau_\Phi,\beta)$-KMS states by the
thermodynamic limit. That is our next topic.

\noindent{\bf The Araki--Gibbs Condition.} For any $\Lambda\in\fG_\mathrm{fin}$,
the so-called surface energies
$$
W_\Lambda(\Phi):=\sum_{\atop{X\cap\Lambda\neq\emptyset}{X\cap\Lambda^\mathrm{c}\neq\emptyset}}\Phi(X)
$$
are self-adjoint elements of $\cO$. Let $\beta>0$ and $V_\Lambda=\beta
W_\Lambda(\Phi)$. Suppose that $\omega$ is a modular state on $\cO$ and let
$\delta_\omega$ be  the generator of its modular $C^\ast$-dynamics
$\varsigma_\omega$.
Consider the perturbed dynamics $\varsigma_{\omega,V_\Lambda}$ generated by
$\delta_\omega+\i[V_\Lambda,\,\cdot\;]$, and let $\omega_{V_\Lambda}$ be the
$(\varsigma_{\omega,V_\Lambda},-1)$-KMS state associated to $\omega$ by Araki's
perturbation theory~\cite[Theorem~5.4.4]{Bratteli1981}. We say that $\omega$
satisfies the $(\beta,\Phi)$ Araki--Gibbs
condition if, for all $\Lambda\in\fG_\mathrm{fin}$, the restriction of
$\omega_{V_\Lambda}$ to $\cO_\Lambda$ is given by~\eqref{thgivs}. The
Araki--Gibbs condition is the quantum extension of the DLR equation in
equilibrium theory of classical spin systems. It has been
introduced in~\cite{Araki1974b}, see also~\cite{Araki1975/76, Bratteli1981,
Simon1993}.

\begin{theorem} Suppose that $\omega$ is a modular state on $\cO$ and $\beta>0$.
Then the following statements are equivalent.
\ben
\item $\omega$ is a $(\tau_\Phi,\beta)$-KMS state.
\item $\omega$ satisfies the $(\beta,\Phi)$ Araki--Gibbs condition.
\een
\end{theorem}

\subsection{The OQ2S setting}
\label{sec-oqss}

Following~\cite{Ruelle2001}, we consider a quantum spin system over a set $G$
with interaction $\Phi$ satisfying:

\begin{quote}{\bf (DEC)} \ben
\item $G$ is the disjoint union
\[
G=S\sqcup\left(\bigsqcup\limits_{j=1}^M R_j\right)
\]
of a finite set $S$ and finitely many countably infinite sets $R_1,\ldots,R_M$.
\item $\fG_\mathrm{fin}$ is the set of finite subsets of $G$, and the indexing set is
$$
\cI=\{\Lambda\in\fG_\mathrm{fin}\mid \Lambda\supset S\}.
$$
\item Besides satisfying Assumption~\SR, the interaction
$\Phi:\fG_\mathrm{fin}\to\cO_G$ is such that $\Phi(X)=0$ whenever there exist
$i\neq j$ with $X\cap R_i\neq\emptyset$ and $X\cap R_j\neq\emptyset$.

\een
\label{dec}
\end{quote}
\newcommand{\DEC}{{\hyperref[dec]{{\rm (DEC)}}}}

Assumption~\DEC{} implies
\[
\cO_G=\cO_S\otimes\left(\bigotimes\limits_{j=1}^M \cO_{R_j}\right),
\]
where $\cO_S$ pertains to the {\sl small systems} $S$, and
$\cO_{R_j}$ to the $j^\mathrm{th}$ {\sl reservoir} $R_j$.

For each $j\in\{1,\ldots,M\}$, we define the interaction
\beq
\Phi_j(X)=\begin{cases}
\Phi(X)&\text{if }X\subseteq R_j;\\[4pt]
0&\text{otherwise,}
\end{cases}
\label{equ:PhijDef}
\eeq
which clearly satisfies $\|\Phi_j\|_\lambda\le\|\Phi\|_\lambda$.
We denote by $\tau_{\Phi_j}$ the associated $C^\ast$-dynamics on
$\cO_{R_j}$ and by $\delta_{\Phi_j}$ its generator. We further define
\beq
V_j=\sum_{\atop{X\subseteq S\cup R_j}{X\cap S\neq\emptyset,X\cap R_j\neq\emptyset}}\Phi(X),\qquad
V=\sum_{j=1}^M V_j,
\label{equ:VDef}
\eeq
which are self-adjoint elements of $\cO$.
The ``free'' $C^\ast$-dynamics $\tau_\fr$ on $\cO_G$ is generated by
\[
\delta_\fr=\delta_S+\sum_{j=1}^M\delta_{\Phi_j},
\]
where $\delta_S=\i[H_S(\Phi), \cdot\;]$. Obviously,
\[
\delta_\Phi =\delta_\fr+\i[V, \cdot\;].
\]
The model is completed by the choice of a $(\tau_{\Phi_j}, \beta_j)$-KMS state $\omega_{\beta_j}$ on $\cO_{R_j}$ for each
$j\in\{1,\ldots,M\}$, and by taking
\beq
\omega=\omega_S\otimes\left(\bigotimes_{j=1}^M\omega_{\beta_j}\right)
\label{equ:omegaProdForm}
\eeq
for the reference state of $(\cO,\tau_\Phi)$, where, for convenience, $\omega_S$ is taken to be the tracial state on $\cO_S$\footnote{None of our results depend on the choice of $\omega_S$ as long as $\omega_S>0$.},
\[
\omega_S(A) =\frac{\tr(A)}{\dim\cK_S}.
\]
Obviously, $(\cO,\tau_\Phi,\omega)$ is an example of open quantum system as discussed in~\cite[Section 1.1]{Benoist2023a}
with reservoirs $R_j$ described by the $C^\ast$-quantum dynamical systems $(\cO_{R_j},\tau_{\Phi_j},\omega_{\beta_j})$.
The modular group of the state $\omega$ is
\beq
\varsigma_\omega^\theta=\tau_{\Phi_1}^{-\beta_1\theta}\circ\cdots\circ\tau_{\Phi_M}^{-\beta_M\theta},
\label{equ:sigmaomegaForm}
\eeq
and its generator is $\delta_\omega=-\sum_j\beta_j\delta_{\Phi_j}$. It follows from the
definitions~\eqref{equ:VDef} that
\beq
\sigma=\delta_\omega(V)=-\sum_{j=1}^M\beta_j\delta_{\Phi_j}(V_j)
=-\sum_{j=1}^M\beta_j\sum_{\atop{X\subseteq S\cup R_j}{X\cap S\not=\emptyset, X\cap R_j\not=\emptyset}}
\delta_{\Phi_j}(\Phi(X)).
\label{equ:sigmaDef}
\eeq
From Theorem~\ref{p-oce}(4) and the definition~\eqref{equ:PhijDef}, we further deduce that
\[
\|\delta_{\Phi_j}(\Phi(X))\|\le
\frac{2}{\lambda}\|\Phi\|_\lambda\e^{\lambda |X|}\|\Phi(X)\|,
\]
so that
$$
\|\sigma\|\le\sum_{j=1}^M\beta_j\sum_{x\in S}\sum_{X\ni x}\|\delta_{\Phi_j}(\Phi(X))\|
\le\frac{2|S|\e^\lambda}{\lambda}\|\Phi\|_\lambda^2\sum_{j=1}^M\beta_j,
$$
and in particular $V\in\Dom(\delta_\omega)$.

We now describe our thermodynamic limit  scheme.
We  write  $\Lambda\in\cI$ as the disjoint union
\[
\Lambda= S\sqcup\left(\bigsqcup_{j=1}^M\Lambda_j\right),\qquad\Lambda_j\in\fG_\mathrm{fin}\cap R_j,
\]
and use the identification
\[
\cO_\Lambda=\cO_S\otimes\left(\bigotimes_{j=1}^M\cO_{\Lambda_j}\right).
\]
Let
\beq
\omega_{\Lambda_ j}=\omega_{\beta_j}\big|_{\cO_{\Lambda_j}},
\label{new-c}
\eeq
and, identifying $\omega_{\Lambda_j}$ with a density matrix on $\cK_{\Lambda_j}$,
\beq
\widehat H_{\fr,\Lambda_j}=-\frac{1}{\beta_j} \log \omega_{\Lambda_j}.
\label{new-c-1}
\eeq
The  finite volume dynamics $\tau_\Lambda$ is generated by the Hamiltonian
\beq
\widehat H_\Lambda=\widehat H_{\fr,\Lambda}+V_\Lambda,
\label{new-c-2}
\eeq
where
\[
\widehat H_{\fr,\Lambda}=H_S(\Phi)+\sum_{j=1}^M\widehat H_{\fr,\Lambda_j},
\]
and
\[
V_\Lambda=\sum_{j=1}^M V_{j,\Lambda},\qquad
V_{j,\Lambda}=\sum_{\atop{X\subseteq S\cup\Lambda_j}{X\cap S\neq\emptyset,X\cap\Lambda_j\neq\emptyset}}\Phi(X).
\]
Finally, we observe that
\beq
\omega_\Lambda=\omega_S\otimes\left(\bigotimes_{j=1}^M\omega_{\Lambda_j}\right)=\omega\big|_{\cO_\Lambda}.
\label{state-nss}
\eeq
The net $(\cO_\Lambda,\tau_\Lambda,\omega_\Lambda)_{\Lambda\in\cI}$ defines our TDL  scheme. The finite volume entropy
production observable is
\beq
\sigma_\Lambda=\delta_{\omega_\Lambda}(V_\Lambda)
=\i[\log\omega_\Lambda,V_\Lambda]=\sum_{j=1}^M\i[\log\omega_{\Lambda_j},V_j].
\label{equ:sigmaLambdaDef}
\eeq
We denote by $(\cH_S,\pi_S,\Omega_S)$ the GNS-representation of $\cO_S$ induced by $\omega_S$\footnote{As in~\cite{Benoist2023a}, we take $\cH_S=\cK_S\otimes\cK_S$, $\pi_S(A)=A\otimes\one$,
$\Omega_S=\frac{1}{\sqrt N}\sum_i\psi_i\otimes\psi_i$, where $\{\psi_i\}$ is an orthonormal basis
of $\cK_S$ and $N=\dim \cK_S$.}
and by $(\cH_j,\pi_j,\Omega_j)$ the GNS-representation of $\cO_{R_j}$ associated to $\omega_{\beta_j}$. For the
GNS-representation of $\cO_G$ associated to $\omega$ we then take $(\cH,\pi,\Omega)$ where
$$
\Omega=\Omega_S\otimes\left(\bigotimes_{j=1}^M\Omega_j\right)\in\cH=\cH_S\otimes\left(\bigotimes_{j=1}^M\cH_j\right),
$$
and
$$
\pi=\pi_S\otimes\left(\bigotimes_{j=1}^M\pi_j\right).
$$
We have a similar product structure for the TDL scheme, where besides $(\cH_S,\pi_S,\Omega_S)$
we take, for $j\in\{1,\ldots,M\}$, $(\cH_{\Lambda_j},\pi_{\Lambda_j},\Omega_{\Lambda_j})$ to be
$$
\pi_{\Lambda_j}=\pi_j\big|_{\cO_{\Lambda_j}},\qquad
\Omega_{\Lambda_j}=\Omega_j,\qquad
\cH_{\Lambda_j}=\pi_j(\cO_{\Lambda_j})\Omega_j\subset \cH_{j}.
$$

Other choices for $\tau_\Lambda$ and $\omega_\Lambda$ are possible. An arguably simpler choice is to take in~\eqref{state-nss}
\beq
\omega_{\Lambda_j}=\frac{\e^{-\beta_j H_{\Lambda_j}(\Phi_j)}}{\tr(\e^{-\beta_j H_{\Lambda_j}(\Phi_j)})},
\label{ru-spin}
\eeq
and for $\tau_\Lambda$ the dynamics $\tau_{\Phi,\Lambda}$ generated by
\beq
H_\Lambda(\Phi)=H_S(\Phi)+\sum_{j=1}^M H_{\Lambda_j}(\Phi_j)+V_\Lambda.
\label{ru-1-spin}
\eeq

However, our proofs do not work for this choice and we will comment more on this point  in Sections~\ref{sec-discussion-spin} and
\ref{sec-remarks-ebbm}.
\subsection{TDL of 2TMEP}
\label{sec-tdl-spin}

Besides Assumption~\DEC{}, we will also need

\begin{quote}{\bf (SE)}  Each reservoir system $(\cO_{R_j}, \tau_{\Phi_j},\omega_{\beta_j})$ is ergodic, {\sl i.e.,} for
any $\omega_{\beta_j}$-normal state $\nu$ on $\cO_{R_j}$ and any $A\in\cO_{R_j}$,
$$
\lim_{T\to\infty}\frac1T\int_0^T\nu\circ\tau_{\Phi_j}^t(A)\d t=\omega_{\beta_j}(A).
$$
\label{se}
\end{quote}
\newcommand{\SE}{{\hyperref[se]{{\rm (SE)}}}}

\SE{} will be used only to verify~\TDLThree(2). We set
\[
\bar\beta_\lambda:=\|\Phi\|_\lambda\max_{j}\beta_j.
\]
\begin{theorem}\label{spin-main-thm-1}
Suppose that~\DEC{} and~\SE{} hold with $\bar\beta_\lambda<\lambda$.
Then Assumptions \TDLOne--\TDLThree{} hold. In particular, Proposition~\ref{sl-exam} and Theorem~\ref{ss-sunday}(1) hold for~{\rm OQ2S}.
\end{theorem}

The rest of this section is devoted to the proof of Theorem~\ref{spin-main-thm-1} through a sequence of
Lemmata.  The road map from \DEC+\SE{} to \TDLOne--\TDLThree{} is summarized in the following chart.

\begin{figure}[h]
\begin{center}
\begin{tabular}{ccccc}
\toprule
\DEC\text{ with }$\lambda>\bar\beta_\lambda$&$\atop{\text{Lemma~\ref{ss-saturday}}}{\Longrightarrow}$&\TDLFive
&$\atop{\text{Proposition~\ref{sd-new}}}{\Longrightarrow}$&\TDLOne\\
\midrule
\DEC&$\atop{\text{Lemma~\ref{igc-new}}}{\Longrightarrow}$&\TDLTwo(1--5)&\TDLSix(1--3)&\\
\midrule
\DEC\text{ with }$\lambda>\bar\beta_\lambda$&$\atop{\text{Lemma~\ref{lemTS}}}{\Longrightarrow}$&\TDLTwo(6)&&\\
\midrule
\DEC\SE&$\atop{\text{Lemma~\ref{lemTE}}}{\Longrightarrow}$&\TDLThree&&\\
\bottomrule
\end{tabular}
\end{center}
\end{figure}

\bel\label{ss-saturday}
Suppose that~\DEC{} holds. Then,
\ben
\item The map
\[
\rr\ni\theta\mapsto\varsigma_{\omega}^\theta(V)\in\cO
\]
has an analytic continuation to the horizontal strip $|\Im\theta|<\frac{\lambda}{2\bar\beta_\lambda}$.
\item  The  map
\[
\rr \ni \theta \mapsto \varsigma_{\omega}^\theta(\sigma)\in \cO
\]
has an analytic continuation to the horizontal strip $|\Im\theta|<\frac{\lambda}{2\bar\beta_\lambda}$.
\item For all $s\in\rr$ the map
\[
\i\rr\ni z\mapsto [D\omega_s:D\omega]_z
\]
has an analytic continuation to the vertical strip $|\Re z|<\frac{\lambda}{2\bar\beta_\lambda}$.
\een
Each of these maps is bounded on any closed substrip of  the respective strip.
In particular, if~\SR{} holds with $\lambda >\bar\beta_\lambda$, then Assumptions~\TDLOne{} and~\TDLFive{} hold.
\eel
\proof {\bf (1)} By Relations~\eqref{equ:VDef} and~\eqref{equ:sigmaomegaForm}, one has
\[
\varsigma_{\omega}^\theta(V)=\sum_{j=1}^M \tau_{\Phi_j}^{-\beta_j\theta}(V_j).
\]
Hence, it suffices to show that for any $j\in\{1,\ldots,M\}$, the function
\beq
\rr\ni t\mapsto\tau_{\Phi_j}^t(V_j)
\label{equ:expotheta}
\eeq
has an analytic continuation to the strip  $|\Im z|<\frac{\lambda}{2\|\Phi\|_\lambda}$.
By Theorem~\ref{p-oce}(4) and the definition~\eqref{equ:PhijDef}, for any positive integer $n$ one has
\begin{align*}
\frac1{n!}\|\delta_{\Phi_j}^n(V_j)\|
&\le\frac1{n!}\sum_{\atop{X\subseteq S\cup R_j}{X\cap S\neq\emptyset,X\cap R_j\neq\emptyset}}\|\delta_{\Phi_j}^n(\Phi(X))\|\\[4pt]
&\le\sum_{x\in S}\sum_{X\ni x}\left(\frac{2\|\Phi\|_\lambda}{\lambda}\right)^n\e^{\lambda |X|} \,\|\Phi(X)\|
\le|S|\e^\lambda\|\Phi\|_\lambda\left(\frac{2\|\Phi\|_\lambda}{\lambda}\right)^n,
\end{align*}
so that the radius of convergence of the Taylor series of the function~\eqref{equ:expotheta} is $\lambda/2\|\Phi\|_\lambda$. The result then follows from the group property of $\tau_{\Phi_j}$.

\medskip\noindent{\bf (2)} Follows from~(1) by differentiation.

\medskip\noindent{\bf (3)} Follows from~(2) and the proof of Proposition~\ref{sd-new}.

It is an immediate consequence of the group property and the isometric nature of $\varsigma_\omega$
that the three maps are uniformly bounded on any closed substrip of their respective strip of analyticity.\hfill \qed

\bel\label{igc-new}
Assumption~\DEC{} implies \TDLTwo{\rm(1--5)} and~\TDLSix{\rm(1--3)}.
\eel
\proof Parts~(1--3) and~(5) of~\TDLTwo{} as well as Part~(1) of~\TDLSix{} are immediate consequences of the definition
of the TDL scheme and do not depend on~\SR{}. The same is true for the first statement of~\TDLSix(2), while~\SR{}
implies the second statement, namely that
\beq
\lim_\Lambda V_\Lambda=V
\label{equ:VlambdaConv}
\eeq
holds in the norm of $\cO$.

\TDLSix(3) is an immediate consequence of Theorem~\ref{araki74}(2). Indeed,
\[
\e^{\i t\cL_\fr}=\prod_{j=1}^M\Delta_{\omega_{\beta_j}}^{-\i\beta_jt},
\]
where $\Delta_{\omega_{\beta_j}}$ is the modular operator of $\omega_j$, and similarly
\[
\e^{\i t\cL_{\Lambda,\fr}}=\prod_{j=1}^M\Delta_{\omega_{\Lambda_j}}^{-\i\beta_jt}.
\]
These two observations and Theorem~\ref{araki74}(2) give that
\beq
\slim_\Lambda\e^{\i t\cL_{\Lambda,\fr}}=\e^{\i t\cL_\fr},
\label{tdlsix-1}
\eeq
locally uniformly for $t\in\rr$.

To prove Part~(4) of~\TDLTwo{}, consider the Dyson expansion
\begin{align*}
&\e^{\i t(\cL_{\Lambda,\fr}+V_\Lambda)}=\e^{\i t\cL_{\Lambda,\fr}}\\[4pt]
&+\sum_{n\ge1}\i^n\int\limits_{0\le t_1\le\cdots\le t_n\le t}
\e^{\i t_1\cL_{\Lambda,\fr}}V_\Lambda\e^{\i (t_2-t_1)\cL_{\Lambda,\fr}}V_\Lambda
\cdots\e^{\i t(t_n-t_{n-1})\cL_{\Lambda,\fr}}V_\Lambda\e^{\i (t-t_n)\cL_{\Lambda,\fr}}
\d t_1\cdots \d t_n.
\end{align*}
Invoking~\eqref{equ:VlambdaConv} and~\eqref{tdlsix-1} gives that
\beq
\slim_{\Lambda}\e^{\i t(\cL_{\Lambda,\fr}+V_\Lambda)}=\e^{\i t(\cL_\fr + V)},
\label{lf-mon}
\eeq
locally uniformly for $t\in\rr$. Finally, for $A\in\cO_\mathrm{loc}$\footnote{One can take here any $A\in\cO_G$.},
\begin{align*}
\slim_\Lambda\tau_{\Lambda}^t(A)
&=\slim_\Lambda\e^{\i t(\cL_{\Lambda,\fr}+V_\Lambda)}A\e^{-\i t(\cL_{\Lambda,\fr}+V_\Lambda)}\\
&=\e^{\i t(\cL_\fr+V)}A\e^{-\i t(\cL_\fr+V)}=\tau^t(A)
\end{align*}
locally uniformly for $t\in\rr$. Note that for this argument it is sufficient that $\slim_{\Lambda}V_\Lambda=V$.\hfill\qed

\bel\label{ss-mon}
Suppose that \DEC{} holds with $\lambda>\bar\beta_\lambda$. Then,
\beq
\slim_\Lambda\varsigma_{\omega_\Lambda}^\theta(V_\Lambda)=\varsigma_\omega^\theta(V),
\label{ax-walk}
\eeq
locally uniformly for $|\Im\theta|\le\frac12$.
\eel
\proof For $\Lambda\in\cI$, let us define the interaction
$\Phi_{\Lambda}$ by
$$
\Phi_{\Lambda}(X)=\begin{cases}
\Phi(X)&\text{if }X\subseteq\Lambda;\\[4pt]
0&\text{otherwise},
\end{cases}
$$
so that
$$
\lim_\Lambda\|\Phi-\Phi_{\Lambda}\|_\lambda
=\lim_\Lambda\sup_{x\in G}\sum_{\atop{X\ni x}{X\cap\Lambda^\mathrm{c}\neq\emptyset}}\|\Phi(X)\|\e^{\lambda(|X|-1)}=0.
$$
Since, for $j\in\{1,\ldots,M\}$,
$$
D_{j,\Lambda}=V_j-V_{j,\Lambda}
=\sum_{\atop{X\subseteq S\cup R_j}{X\cap S\neq\emptyset\neq X\cap\Lambda_j^\mathrm{c}}}\Phi(X),
$$
the estimate of Theorem~\ref{p-oce}(4) gives
$$
\frac1{n!}\|\delta_{\Phi_j}^n(D_\Lambda)\|
\le|S|\e^\lambda\left(\frac{2\|\Phi\|_\lambda}{\lambda}\right)^n\sup_{x\in S}
\sum_{\atop{X\ni x}{X\cap\Lambda^\mathrm{c}\neq\emptyset}}
\|\Phi(X)\|\e^{\lambda(|X|-1)}
\le|S|\e^\lambda\left(\frac{2\|\Phi\|_\lambda}{\lambda}\right)^n\|\Phi-\Phi_\Lambda\|_\lambda,
$$
which, by the argument in the proof of Lemma~\ref{ss-saturday}, implies that
\beq
\lim_\Lambda\varsigma_\omega^\theta(V_\Lambda)=\varsigma_\omega^\theta(V)
\label{equ:sigomconv}
\eeq
in norm, locally uniformly for $|\Im\theta|<\frac{\lambda}{2\bar\beta_\lambda}$.

Since $\Omega_\Lambda=\Omega$ and $\pi_\Lambda=\pi$, for any $A\in\cO_\Lambda$, the definition of
the modular operator and conjugation gives
\[
J_\Lambda\Delta_{\omega_\Lambda}^{\frac12}\pi(A)\Omega=J\Delta_{\omega}^{\frac12}\pi(A)\Omega.
\]
Since $\pi(\cO_\Lambda)\Omega=\cH_\Lambda$, we have that, as operators on $\cH_\Lambda$,
$$
\Delta_{\omega_\Lambda}^{\frac{1}{2}}= J_\Lambda J \Delta_{\omega}^{\frac{1}{2}},\qquad
\Delta_{\omega_\Lambda}^{-\frac{1}{2}}= \Delta_{\omega}^{-\frac{1}{2}}JJ_\Lambda,
$$
and so
\beq
\begin{split}
\varsigma_{\omega_\Lambda}^\frac12(V_\Lambda)=\Delta_{\omega_\Lambda}^{\frac{1}{2}}V_\Lambda \Delta_{\omega_\Lambda}^{-\frac{1}{2}}&= J_\Lambda J \Delta_{\omega}^{\frac{1}{2}}V_\Lambda
\Delta_{\omega}^{-\frac{1}{2}}JJ_\Lambda,\\[1mm]
\varsigma_{\omega_\Lambda}^{-\frac12}(V_\Lambda)=
\Delta_{\omega_\Lambda}^{-\frac{1}{2}}V_\Lambda \Delta_{\omega_\Lambda}^{\frac{1}{2}}&= J_\Lambda J \Delta_{\omega}^{-\frac{1}{2}}V_\Lambda
\Delta_{\omega}^{\frac{1}{2}}JJ_\Lambda,
\end{split}
\label{icn-am}
\eeq
on $\cH_\Lambda$. By Theorem~\ref{araki74}(2--3), the relations~\eqref{equ:sigomconv}--\eqref{icn-am} and the group
property give that
\[
\slim_\Lambda\varsigma_{\omega_\Lambda}^\theta(V_\Lambda)
=\slim_\Lambda\Delta_{\omega_\Lambda}^{\i\Re\theta}J_\Lambda J\varsigma_\omega^{\i\Im\theta}(V_\Lambda)JJ_\Lambda\Delta_{\omega_\Lambda}^{-\i\Re\theta}
=\varsigma_\omega^\theta(V),
\]
locally uniformly for $|\Im\theta|=\frac12$.

On the one hand, considering the functions
$$
f_\Lambda(\theta)=\e^{-\theta^2}(\varsigma_\omega^\theta(V)-\varsigma_{\omega_\Lambda}^\theta(V_\Lambda))\Psi
$$
with $\Psi\in\cH$, this gives that, for any $\epsilon>0$ and $L>0$, there is $\Lambda_{\epsilon,L}$ such that,
for $\Lambda\supset\Lambda_{\epsilon,L}$,
$$
\sup_{\theta\in[-L,L]}\|f_\Lambda(\theta\pm\tfrac\i2)\|<\epsilon.
$$
On the other hand, for $\theta\in\R$, there exists a constant $C$ such that
$$
\|\varsigma_\omega^{\theta\pm\frac\i2}(V)\Psi\|\le\|\varsigma_\omega^{\pm\frac\i2}(V)\|\|\Psi\|\le\tfrac C6\e^{\frac14},
$$
and, invoking~\eqref{icn-am}+\eqref{equ:sigomconv} and making $\Lambda_{\epsilon,L}$ possibly larger,
$$
\|\varsigma_{\omega_\Lambda}^{\theta\pm\frac\i2}(V_\Lambda)\Psi\|
\le\|\varsigma_{\omega_\Lambda}^{\pm\frac\i2}(V_\Lambda)\|\|\Psi\|
=\|\varsigma_{\omega}^{\pm\frac\i2}(V_\Lambda)\|\|\Psi\|
\le 2\|\varsigma_\omega^{\pm\frac\i2}(V)\|\|\Psi\|\le\tfrac {2C}6\e^{\frac14},
$$
provided $\Lambda\supset\Lambda_{\epsilon,L}$. It follows that
$$
\sup_{\theta\in\R\setminus[-L,L]}\|f_\Lambda(\theta\pm\tfrac\i2)\|\le C\e^{-L^2},
$$
and hence, for $0<\epsilon<C$ and choosing $L>\sqrt{\log\frac C\epsilon}$, we derive that
$$
\sup_{\theta\in\R}\|f_\Lambda(\theta\pm\tfrac\i2)\|<\epsilon
$$
provided $\Lambda$ is large enough. Applying Hadamard's three lines theorem gives that
$$
\sup_{|\Im\theta|\le\frac12}\|f_\Lambda(\theta)\|<\epsilon
$$
for $\Lambda$ is large enough. This yields the convergence~\eqref{ax-walk} with the required uniformity.
\hfill\qed

\bel\label{lemTS}
Suppose that~\DEC{} holds with $\lambda>\bar\beta_\lambda$. Then~\TDLTwo{\rm(6)} holds.
\eel
\proof By Relations~\eqref{equ:sigmaDef} and~\eqref{equ:sigmaLambdaDef} and Lemma~\ref{ss-saturday}(1), we have
$$
\partial_\theta\varsigma_{\omega_\Lambda}^\theta(V_\Lambda)=\varsigma_{\omega_\Lambda}^\theta(\sigma_\Lambda),
\qquad
\partial_\theta\varsigma_\omega^\theta(V)=\varsigma_\omega^\theta(\sigma),
$$
for $|\Im\theta|<\frac12$. Thus, Lemma~\ref{ss-mon} and Weierstrass' convergence theorem allow us to conclude that
\beq
\slim_{\Lambda}\varsigma_{\omega_\Lambda}^\theta(\sigma_\Lambda)=\varsigma_\omega^\theta(\sigma),
\label{ax-walk-1}
\eeq
locally uniformly for $|\Im\theta|<\frac12$. In particular,
\[
\slim_{\Lambda}\sigma_\Lambda=\sigma,
\]
and, invoking the uniform boundedness principle, $\sup_\Lambda \|\sigma_\Lambda\|<\infty$.
Hence, using~\eqref{lf-mon}, we get,
\beq
\slim_{\Lambda}\int_0^s \tau_\Lambda^{-t}(\sigma_\Lambda) \d t=\int_0^s \tau^{-t}(\sigma)\d t.
\label{equ:EPconv}
\eeq
By Lemma~\ref{lem:relDeltaForm}, we further have
\begin{align*}
\log\Delta_{\omega_{\Lambda,s}|\omega_\Lambda}
&=\log\Delta_{\omega_\Lambda}+\int_0^s\tau_\Lambda^{-t}(\sigma_\Lambda)\d t,\\[4pt]
\log\Delta_{\omega_s|\omega}
&=\log\Delta_\omega+\int_0^s\tau^{-t}(\sigma)\d t,
\end{align*}
for all $s\in\rr$. By Theorem~\ref{araki74}(2) and Trotter's theorem~\cite[Theorem~VIII.21]{Reed1980}, these two
relations and~\eqref{equ:EPconv} allow us to conclude that, for all $s\in\rr$,
\[
\lim_\Lambda\log\Delta_{\omega_{\Lambda,s}|\omega_\Lambda}=\log\Delta_{\omega_s|\omega}
\]
holds in the strong resolvent sense. By Trotter's theorem again
$$
\slim_\Lambda\Delta_{\omega_{\Lambda,s}|\omega_\Lambda}^{\i\theta}=\Delta_{\omega_s|\omega}^{\i\theta},
$$
holds for all $\theta,s\in\rr$, and together with Theorem~\ref{araki74}(2),
$$
\slim_\Lambda\Delta_{\omega_{\Lambda,s}|\omega_\Lambda}^{\i\theta}\Delta_{\omega_\Lambda}^{-\i\theta}
=\Delta_{\omega_s|\omega}^{\i\theta}\Delta_{\omega}^{-\i\theta},
$$
which is~\TDLTwo(6). \hfill\qed

Finally, we complete the proof of Theorem~\ref{spin-main-thm-1} with
\bel\label{lemTE}
Assumption~\SE{} implies~\TDLThree.
\eel
\proof \TDLThree(1) follows from Theorem~\ref{araki74}(2). Given the product structure~\eqref{equ:omegaProdForm} of the
reference state $\omega$, one has
\beq
\log\Delta_\omega=\sum_{j=1}^M\log\Delta_{\omega_{\beta_j}}\
=-\sum_{j=1}^M\beta_j\cL_j,
\label{mar-lat-1}
\eeq
where $\cL_j$ is the standard Liouvillean of the $C^\ast$-dynamics
$\tau_{\Phi_j}$. It was proven in~\cite[Theorem~1.2]{Jaksic2001b} that the $C^\ast$ dynamical system
$(\cO_{R_j},\tau_{\Phi_j},\omega_{\beta_j})$ is ergodic iff  $0$ is a simple eigenvalue of ${\cal L}_j$ and ${\cal L}_j$ has no other
eigenvalues\footnote{For this result it is essential that $\omega_{\beta_j}$ is $(\tau_{\Phi_j}, \beta_j)$-KMS state.}, and so Assumption~\SE{} gives that\footnote{Note that the factors in the sum \eqref{mar-lat-1} commute.}
\[
\Ker \log \Delta_\omega=\cH_S\otimes\Omega_1\otimes \cdots\otimes \Omega_M.
\]
By the definition~\eqref{state-nss} of $\omega_\Lambda$,
$$
\log\Delta_{\omega_\Lambda}=\sum_{j=1}^M\log\Delta_{\omega_{\Lambda_j}},
$$
and since $\Ker\log\Delta_{\omega_{\Lambda_j}}\ni\Omega_{\Lambda_j}=\Omega_j$, one has
\[
\cH_S\otimes \Omega_1\otimes \cdots\otimes \Omega_M\subseteq \Ker \log \Delta_{\omega_\Lambda}.
\]
This gives that~\TDLThree(2) follows from~\SE{} . \hfill\qed
\subsection{Remarks}
\label{sec-discussion-spin}

{\bf 1. On the choice of $\omega_\Lambda$ and $\tau_\Lambda$.} The  proof of
Theorem~\ref{spin-main-thm-1} combines fairly standard thermodynamic limit
arguments with Araki's continuity Theorem~\ref{araki74}. The
definitions~\eqref{new-c}--\eqref{new-c-2} of $\omega_\Lambda$ and $\tau_\Lambda$
take into the account the interaction ``boundary terms'' in a way that ensures the necessary
continuity of the modular structure that is behind our TDL scheme.  This
continuity is broken with choice~\eqref{ru-spin} and~\eqref{ru-1-spin}, and in
this case it is likely that a completely new strategy of the proof is needed.

We remark  that for the choice~\eqref{ru-spin}--\eqref{ru-1-spin} it is not
difficult to  show  that for all $t\in\rr$ and $\alpha\in\i\rr$,
\beq
 \lim_\Lambda\varsigma_{\omega_\Lambda}^\theta([D\omega_{\Lambda,-t}:D\omega_\Lambda]_{\alpha})
 =\varsigma_\omega^\theta([D\omega_{-t}:D\omega]_{\alpha})
\label{ncw}
 \eeq
in norm, locally uniformly for $\theta\in\rr$. Hence, if $\omega$ is a
thermodynamic limit KMS-state and $\omega_{\Lambda_\gamma}$ is subnet such that
$\lim_{\Lambda_\gamma}\bar\omega_{\Lambda_\gamma}=\omega$, we have
\[
\lim_{\Lambda_\gamma} \varsigma_{\omega_{\Lambda_\gamma}}^\theta([D\omega_{\Lambda_\gamma, -t}: D\omega_{\Lambda_\gamma}]_{\alpha})=\varsigma_\omega^\theta
([D\omega_{-t}: D\omega]_{\alpha}),
\]
providing the desired thermodynamic justification
of~\eqref{emm-new}--\eqref{han-ajde-new} in the case
$\nu=\omega$ that is much simpler than the proof
of Theorem~\ref{spin-main-thm-1}. The
convergence~\eqref{ncw} also gives that
\[
\lim_{R \rightarrow \infty}\lim_\Lambda
\frac{1}{R}\int_0^R \nu_\Lambda\left(\varsigma_{\omega_\Lambda}^\theta\left([D\omega_{\Lambda, -t}: D\omega_\Lambda]_{\alpha}\right)\right) \d\theta=
 \lim_{R \rightarrow \infty}\frac{1}{R}\int_0^R
 \nu\left(\varsigma_\omega^\theta\left([D\omega_{-t}: D\omega]_{\alpha}\right)\right) \d\theta
\]
if $\nu_\Lambda \rightarrow \nu$ weakly. Thus, the TDL justification
of~\eqref{emm-new}--\eqref{han-ajde-new} is equivalent to the
justification of the exchange limits
\[
\lim_{R \rightarrow \infty}\lim_\Lambda
\frac{1}{R}\int_0^R \nu_\Lambda\left(\varsigma_{\omega_\Lambda}^\theta\left([D\omega_{\Lambda, -t}: D\omega_\Lambda]_{\alpha}\right)\right) \d\theta=
\lim_\Lambda\lim_{R \rightarrow \infty}
\frac{1}{R}\int_0^R \nu_\Lambda\left(\varsigma_{\omega_\Lambda}^\theta\left([D\omega_{\Lambda, -t}: D\omega_\Lambda]_{\alpha}\right)\right) \d\theta.
\]
It remains an open question whether this holds for the
choice~\eqref{ru-spin}--\eqref{ru-1-spin}. For the
choice~\eqref{new-c}--\eqref{new-c-1}
and the special class of $\nu$'s this question is resolved in
Theorem~\ref{spin-main-thm-1} by a strategy that invokes
Araki's continuity results and the reservoir ergodicity assumptions.

{\bf 2. On the ergodicity assumption.}  The reservoir ergodicity assumption
plays an important role in the 2TMEP stability result established
in~\cite[Theorem 1.5]{Benoist2023a}. However, its  emergence in the study of TDL
of 2TMEP perhaps comes as a surprise.  Although the need for \SE{} is clear
given our strategy of the proof, its  role in the TDL of 2TMEP remains to be
understood better. Needless to say, although \SE{} is believed to hold for
generic spin interactions, there are very few examples for which it has been
established. The case of EBBM is very different and there  the respective
reservoir ergodicity assumption follows from a simple natural criterion.

{\bf 3. On the assumption $\lambda >\bar \beta_\lambda$.}  This
high-temperature assumption also appeared in the  recent work~\cite[Theorem
2.11]{Jaksic2023a} and ensures that the reservoirs states $\omega_{\beta_j}$ are
weak Gibbs, although we did not make use of this fact. The problem of TDL
justification of the 2TMEP for OQ2S for arbitrary $\beta_j$'s remains open.

{\bf 4. TDL of the entropy balance equation.} We recall that the relative
entropy of two density matrices $\rho$ and $\nu$ is
\beq
\mathrm{Ent}(\nu|\rho)=\tr(\nu(\log\rho-\log\nu)).
\label{rel-ent-1}
\eeq
Its basic property is that $\mathrm{Ent}(\nu|\rho)\le0$ with equality iff $\rho=\nu$.

In the general setting of algebraic statistical mechanics, the relative entropy
of a pair $(\nu,\rho)$ of normal states has been introduced by Araki in the
seminal papers~\cite{Araki1975/76,Araki1977}. With the sign and ordering
convention of~\cite{Jaksic2001a}, Araki's definition reads
\[
\mathrm{Ent}(\nu|\rho)=\langle\Omega_\nu,\log\Delta_{\rho|\nu}\Omega_\nu\rangle.
\]
It shares the above mentioned basic property with~\eqref{rel-ent-1} and reduces
to~\eqref{rel-ent-1} in the finite dimensional setting.

For additional information about relative entropy we refer the reader to~\cite{Ohya1993}.

Returning to our OQ2S, Lemma~\ref{lem:relDeltaForm} gives that
\beq
\mathrm{Ent}(\omega_s|\omega)=-\int_0^s\omega_t(\sigma)\d t,
\label{bala-1}
\eeq
and similarly, for either of the choices~\eqref{new-c}--\eqref{new-c-2}/\eqref{ru-spin}--\eqref{ru-1-spin},
\beq
\mathrm{Ent}(\omega_{\Lambda,s}|\omega_\Lambda)=-\int_0^s\omega_{\Lambda,t}(\sigma_\Lambda)\d t,
\label{bala-2}
\eeq
with $\sigma_\Lambda=\delta_{\omega_\Lambda}(V_\Lambda)$. The
identities~\eqref{bala-1}--\eqref{bala-2} are the entropy balance
equations of~\cite{Ruelle2001,Jaksic2001a}. Note that for the
choice~\eqref{ru-spin}--\eqref{ru-1-spin},
\beq
\sigma_\Lambda=\delta_{\omega_\Lambda}(V_\Lambda)
=-\sum_{j=1}^M\beta_j\,\i[H_{\Lambda_j}(\Phi_j),V_\Lambda]
=\sum_{j=1}^M\beta_j\,\i[H_\Lambda(\Phi), H_{\Lambda_j}(\Phi_j)],
\label{rue-obs}
\eeq
which is Ruelle's original definition of the entropy production observable in~\cite{Ruelle2001}.

For the choice~\eqref{ru-spin}--\eqref{ru-1-spin}, assuming~\SR, Theorem~\ref{p-oce}(1+4) ensures that
for $A\in\cO_\mathrm{loc}$ and locally uniformly for $t\in\rr$,
$$
\lim_\Lambda\sigma_\Lambda=\sigma,\qquad
\lim_\Lambda\tau_\Lambda^t(A)=\tau^t(A),
$$
hold in norm. Given $\epsilon>0$, there is $\Lambda'\in\cI$ such that $\|\sigma-\sigma_{\Lambda'}\|<\epsilon$.
Let $\bar\omega_\Lambda$ be an arbitrary extension of $\omega_\Lambda$ to a state on $\cO$ such that $\omega$ is the weak$^\ast$-limit of the subnet $\bar\omega_{\Lambda_\gamma}$. Then, we have
\begin{align*}
|(\omega_t-\bar\omega_{\Lambda_\gamma,t})(\sigma)|&\le
|(\omega_t-\bar\omega_{\Lambda_\gamma,t})(\sigma_{\Lambda'})|
+|(\omega_t-\bar\omega_{\Lambda_\gamma,t})(\sigma-\sigma_{\Lambda'})|\\[4pt]
&\le|(\omega_t-\bar\omega_{\Lambda_\gamma,t})(\sigma_{\Lambda'})|+2\epsilon\\[4pt]
&\le|(\omega-\bar\omega_{\Lambda_\gamma})\circ\tau^t(\sigma_{\Lambda'})|
+|\bar\omega_{\Lambda_\gamma}\circ(\tau^t-\tau_{\Lambda_\gamma}^t)(\sigma_{\Lambda'})|+2\epsilon,
\end{align*}
so that
$$
\limsup_{\Lambda_\gamma}|(\omega_t-\bar\omega_{\Lambda_\gamma,t})(\sigma)|\le2\epsilon,
$$
from which we can conclude that $\lim_{\Lambda_\gamma}\bar\omega_{\Lambda_\gamma,t}(\sigma)=\omega_t(\sigma)$
and hence that
$$
\lim_{\Lambda_\gamma}|\omega_t(\sigma)-\omega_{\Lambda_\gamma,t}(\sigma_{\Lambda_\gamma})|
\le\lim_{\Lambda_\gamma}|\omega_t(\sigma)-\bar\omega_{\Lambda_\gamma,t}(\sigma)|
+\lim_{\Lambda_\gamma}|\bar\omega_{\Lambda_\gamma,t}(\sigma-\sigma_{\Lambda_\gamma})|
=0.
$$
Thus, \eqref{bala-1}-\eqref{bala-2} give
\[
\lim_{\Lambda_\gamma}\mathrm{Ent}(\omega_{\Lambda_\gamma, s}|\omega_{\Lambda_\gamma})
=\mathrm{Ent}(\omega_{s}|\omega),
\]
for all $s\in\rr$. Similarly, for the choice~\eqref{new-c}--\eqref{new-c-2}
and assuming that~\SR{} holds with $\lambda>\bar\beta_\lambda$,
\[
\lim_{\Lambda}{\rm Ent}(\omega_{\Lambda, s}|\omega_\Lambda)= {\rm Ent}(\omega_{s}|\omega)
\]
for all $s\in\rr$.

To connect further with the work~\cite{Ruelle2001}, one can consider an intermediate scenario
where the state $\omega_\Lambda$ is chosen by
\eqref{new-c} and $\tau_\Lambda$ and $\sigma_\Lambda$ by~\eqref{ru-1-spin} and~\eqref{rue-obs}. The average entropy production over the time interval
$[0, s]$ is then
\[-\int_0^s \omega_\Lambda(\tau_\Lambda^t(\sigma_\Lambda))\d t.\]
This entropy production cannot be directly linked to the relative entropy and {does not  need} to be non-negative. However, since
\[
-\lim_\Lambda \int_0^s \omega_\Lambda(\tau_\Lambda^t(\sigma_\Lambda))\d t
=-\int_0^s \omega_t(\sigma)\d t
=\Ent(\omega_s|\omega),
\]
 the basic properties  of the averaged entropy production are restored in the thermodynamic limit. Note that for this choice our TDL proofs for 2TMEP
  do not work due to the lack of commutativity of  the groups $\tau_{\Lambda_j, \fr}$ and $\varsigma_{\omega_{\Lambda_j}}$. Attempts to control this non-commutativity fail
  for  reasons we will discuss  in the next and final remark.

In summary, the TDL of the averaged entropy production of OQ2S is robust and does not exhibit the same subtleties as the TDL of 2TMEP.


{\bf 5. On the Araki--Gibbs Condition.} The reader may have noticed that our
failure to establish \TDLSix(4) stems from the restriction
$|\Im\theta|\le\frac12$ in Lemma~\ref{ss-mon} that is forced by the
identities~\eqref{icn-am}. This restriction allows us to
establish~\eqref{ax-walk-1} only for $|\Im\theta|<\frac12$ and we cannot
reach~\TDLSix(4) due to the lack of control on the line $\Im\theta=\frac12$.
One can attempt a different approach  by using directly~\eqref{icn-am} with
$\sigma_\Lambda$ instead of $V_\Lambda$ or by using Theorem~\ref{araki74}(4),
and indeed for the choice~\eqref{ru-spin}--\eqref{ru-1-spin} where
$\sigma_\Lambda$ is given by~\eqref{rue-obs} this works relatively effortlessly.
However, for the choice~\eqref{new-c}--\eqref{new-c-1} one faces perennial
difficulties in controlling $\log \omega_\Lambda$ on the basis of the
Araki--Gibbs Condition; see~\cite[Section 2.2]{Jaksic2023a} for related
discussion and references. Comparing with Remark 1, one arrives at the following
dual problems that must be understood better before  further progress is made:

(a) The lack of continuity of modular structure, in the spirit of Theorem~\ref{araki74},  for the choice~\eqref{ru-spin}--\eqref{ru-1-spin}.

(b) The lack of control of $\log\omega_\Lambda$ and $\sigma_\Lambda$, on the basis of the Araki--Gibbs condition, for the choice~\eqref{new-c}--\eqref{new-c-1}.

These problems are absent in the EBBM and related models that involve non-interacting ideal reservoirs.

\section{The Electronic Black Box Model}
\label{sec-main-ebbb}

\subsection{The free Fermi gas}
\label{sec-free-fermi}
Let $\fh$ and $h$ be the single fermion Hilbert space and
Hamiltonian. We denote by $\CAR(\fh)$ the CAR algebra over
$\fh$, and  by $a^\ast(f)/a(f)/\varphi(f)$  the
creation/annihilation/field operator associated to $f\in\fh$. $a^\#$ stands
for either $a$ or $a^\ast$, and we recall that $\|a^\#(f)\|=\|f\|$. The group
$\tau$ of Bogoliubov $\ast$-automorphisms of $\CAR(\fh)$ generated by $h$
is uniquely specified by $\tau^t(a^\#(f))=a^\#(\e^{\i th}f)$. The
$C^\ast$-dynamical system $(\CAR(\fh),\tau)$ describes the free Fermi gas over
$(\fh,h)$. We denote by $\CAR_g(\fh)$ the gauge-invariant
$C^\ast$-subalgebra of $\CAR(\fh)$ generated by
$\{a^\ast(f)a(g)\mid f,g\in\fh\}\cup\{\one\}$.
The dynamics $\tau$ obviously preserves $\CAR_g(\fh)$.
For a given distribution operator $0\le T<\one$ on $\fh$, $\omega_T$
denotes the quasi-free state on $\CAR(\fh)$ generated by $T$. The same letter
will be used for the  restriction of $\omega_T$ to $\CAR_g(\fh)$.
$\omega_T$ is faithful iff $\Ker T$ is trivial, which we assume to hold in
what follows. Of particular importance is the Fermi-Dirac distribution
\[
T_{\beta\mu}=\frac{1}{1+\e^{\beta(h-\mu\one)}},
\]
where $\beta>0$ and $\mu\in\rr$. We write $\omega_{\beta\mu}$ for
$\omega_{T_{\beta\mu}}$ and remark that $\omega_{\beta\mu}$ is a
$(\tau,\beta)$-KMS state on $\CAR_g(\fh)$. The $C^\ast$-quantum dynamical system
$(\CAR(\fh),\tau,\omega_{\beta\mu})$ describes a free Fermi gas in thermal
equilibrium at inverse temperature $\beta$ and chemical potential $\mu$. The
subsystem $(\CAR_g(\fh),\tau,\omega_{\beta\mu})$ describes its
gauge-invariant part, and will be the focus of this section. The full system
$(\CAR(\fh),\tau,\omega_{\beta\mu})$ will be discussed in Section~\ref{sec-remarks-ebbm}.

We recall the following well-known fact, see~\cite{Pillet2006} for a
pedagogical discussion.
\begin{theorem}
Suppose that $h$ commutes with $T$ and has purely absolutely continuous spectrum. Then the quantum
dynamical systems $(\CAR(\fh),\tau,\omega_T)$ and $(\CAR_g(\fh),\tau,\omega_T)$ are mixing,
and in particular ergodic.
\end{theorem}

We will make use of the Araki--Wyss GNS-representation
$(\cH_\mathrm{AW},\pi_\mathrm{AW},\Omega_\mathrm{AW})$ of
$\CAR(\fh)$ associated to $\omega_T$. This representation was
constructed in~\cite{Araki1964a}; see~\cite{Derezinski2013,Jaksic2010b}
for pedagogical introductions to the topic. We assume that  the reader
is familiar with the fermionic second quantization, and for definiteness
we will use the same notation as in~\cite{Jaksic2002a}. We fix a complex
conjugation $\overline{\,\cdot\,}$ on $\fh$ and assume that it  commutes
with $T$\footnote{In the case of $T_{\beta\mu}$ this is the same as
assuming that it commutes with $h$} and $h$. The Araki--Wyss representation
is given by
\begin{align*}
\cH_\mathrm{AW}&=\Gamma_-(\fh)\otimes\Gamma_-(\fh),\\[4pt]
\pi_\mathrm{AW}(a(f))&=a((\one-T)^{1/2}f)\otimes\one+\vartheta\otimes a^\ast(T^{1/2}\bar f),\\[4pt]
\pi_\mathrm{AW}(a^\ast(f))&=a^\ast((\one-T)^{1/2}f)\otimes\one+\vartheta\otimes a(T^{1/2}\bar f),\\[4pt]
\Omega_\mathrm{AW}&=\Omega_\mathrm{f}\otimes\Omega_\mathrm{f},
\end{align*}
where $\Gamma_-(\fh)$ is the fermionic Fock space over $\fh$,
$\vartheta=\Gamma(-\one)=\e^{\i\pi N}$ where $N$ is the number operator, and
$\Omega_\mathrm{f}$ is the Fock vacuum on $\Gamma_-(\fh)$. The standard
Liouvillean of $\tau$ is
\[
\cL=\d\Gamma(h)\otimes\one-\one\otimes\d\Gamma(h),
\]
and
\[
\log\Delta_{\omega_T}=\d\Gamma(k_T)\otimes\one-\one\otimes\d\Gamma(k_T),
\]
where $k_T=\log (T(\one-T)^{-1})$. Note that if $T=T_{\beta\mu}$, then
$k_T=-\beta(h-\mu\one)$ and
\[
\Delta_{\omega_{\beta\mu}}=\e^{-\beta(\cL-\mu {\cal N})},
\]
where ${\cal N}=N\otimes\one -\one \otimes N$.  The chosen complex conjugation on $\fh$ naturally extends to $\Gamma_-(h)$ and
we denote it by the same symbol $\Psi\mapsto\bar\Psi$. The modular conjugation
acts as
\beq
J(\Phi\otimes\Psi)=u\bar\Psi\otimes u\bar\Phi,
\label{aw-mod-con}
\eeq
where $u=\e^{\i\pi N(N-\one)/2}$.

The Araki--Wyss representation of $\CAR_g(\fh)$ associated to $\omega_T$ is
obtained by the obvious restriction.

\subsection{The EBBM setting}
\label{sec-ebmm-setting}
Let $G$ be a countably infinite set satisfying Assumption~\DEC(1) of Section \ref{sec-oqss}. Let  $h$ be a
bounded self-adjoint operator on the Hilbert space $\ell^2(G)$. We denote by
$(\delta_x)_{x\in G}$ the standard basis of $\ell^2(G)$ and set
$h(x,y)=\langle\delta_x,h\delta_y\rangle$.\footnote{We use the corresponding
notation for matrix elements of bounded operators on $\ell^2(G)$.}
Obviously,
\[
\ell^2(G)=\ell^2 (S)\oplus\left(\bigoplus_{j=1}^M\ell^2(R_j)\right).
\]
Let $h_S$ and $h_j$ be the compressions of $h$ to $\ell^2(S)$ and
$\ell^2(R_j)$\footnote{Whenever the meaning is clear within the context,
we denote with the same letters the extensions of these operators to
$\ell^2(G)$ by setting them to zero on $\ell^2(S)^\perp$ and
$\ell^2(R_j)^\perp$. Analogous  extensions will be used without
further saying.}, and
\[
h_\fr=h_S+\left(\bigoplus_{j=1}^Mh_j\right).
\]
The $C^\ast$-algebra of the EBBM is $\cO=\CAR_g(\ell^2(G))$ and its free
$C^\ast$-dynamics $\tau_\fr$ is the group of Bogoliubov $\ast$-automorphisms
generated by $h_{\fr}$. The $C^\ast$-algebra of the small system $S$ and the
$j$-th reservoir $R_j$  are $\cO_S=\CAR_g(\ell^2(S))$  and
$\cO_{R_j}=\CAR_g(\ell^2(R_j))$. Their dynamics $\tau_S$ and $\tau_{j}$
are the groups of Bogoliubov $\ast$-automorphisms generated by $h_S$ and $h_j$.

We write $h=h_\fr+v$ and, besides~\DEC(1), we assume
\begin{quote}{\bf (DEC)}
\ben
\setcounter{enumi}{3}
\item No direct coupling between reservoirs
$$
(x,y)\in R_i\times R_j\text{ for }i\neq j\quad\Longrightarrow\quad v(x,y)=0.
$$

\item $v$ is finite rank, {\sl i.e.,} for some finite $G_0\subset G$ containing $S$,
$$
(x,y)\notin G_0\times G_0\quad\Longrightarrow\quad v(x,y)=0.
$$
\item $\fG_\mathrm{fin}$ is the set of finite subsets of $G$ and the indexing set is
$$
\cI=\{\Lambda\in\fG_\mathrm{fin}\mid\Lambda\supset G_0\}.
$$
\een
\label{dec2}
\end{quote}
\newcommand{\DECtwo}{{\hyperref[dec2]{{\rm (DEC)}}}}
Note that $v=\sum_{j=1}^M v_j$, where $v_j$ denotes the
compression of $v$ to $\ell^2(S\cup R_j)$. The coupling
between $S$ and $R_j$ is given by the ``hopping'' term
\[
V_j=\d\Gamma(v_j)=\sum_{{x,y}\in S\cup R_j} v_j(x,y)a_x^\ast a_y,
\]
where $a_x^\#=a^\#(\delta_x)$. Note that since $v_j$ is finite
rank, $V_j\in\CAR(\ell^2(S\cup R_j))$.

Finally, the self-interaction, restricted to the small system $S$, is described by a self-adjoint element $W$
of  $\CAR_g(\ell^2(S))$. The interacting $C^\ast$-dynamics $\tau$ on $\cO$ is generated by
$\delta=\delta_\fr +\i[V, \cdot\;]$ where
$$
V=W+\sum_{j=1}^MV_j
$$
is a self-adjoint element of $\cO$.

The reference state of the coupled system is the quasi-free state $\omega$ generated by
\[
T=\one_S\oplus\left(\bigoplus_{j=1}^MT_{\beta_j\mu_j}\right),
\]
where $\beta_j>0$ and $\mu_j\in \rr$ are the inverse temperature and chemical potential of the $R_j$ and\footnote{We denote by $\one_S$, $\one_{R_j}$, etc, the orthogonal projection onto $\ell^2(S)$, $\ell^2(R_j)$, etc.}
$$
T_{\beta_j\mu_j}=\frac1{\one+\e^{\beta_j(h_j-\mu_j\one_{R_j})}}
$$
is the corresponding density operator on $\ell^2(R_j)$. The EBBM is described by the quantum
dynamical system $(\cO,\tau,\omega)$. Note that
$\omega\big|_{\cO_{R_j}}=\omega_{\beta_j\mu_j}$ is the $(\tau_j,\beta_j)$-KMS state on $\cO_{R_j}$.

Due to the fermionic statistics,  the open quantum  system $S+R_1+\cdots+R_M$
does not have the tensor product structure with respect to
its  subsystems, as postulated in~\cite[Section 1.1]{Benoist2023a}.
This however does not affect any of the results of~\cite{Benoist2023a},
the proof of Theorem~1.5 in~\cite{Benoist2023a} requiring only notational changes.
We note in particular that disjointedly supported elements of $\cO$ commute, {\sl i.e.,}
if $A\in\cO_T$, $B\in\cO_{T'}$ with $T,T'\subset G$ such that $T\cap T'=\emptyset$, then $[A,B]=0$.

We now describe the TDL approximation scheme.
We  start with  $\Lambda\in\cI$ and write it as
\[
\Lambda=S\sqcup\left(\bigsqcup_{j=1}^M\Lambda_j\right),
\]
where $\Lambda_j\subset R_j$. Let
\[
\cO_{\Lambda_j}= \CAR_g(\ell^2(\Lambda_j)),\qquad
\cO_\Lambda=\CAR_g(\ell^2(\Lambda)).
\]
The free dynamics $\tau_{\fr,\Lambda}$ is the group of Bogoliubov $\ast$-auto\-morphisms generated by
\[
h_{\fr,\Lambda}= h_S\oplus\left(\bigoplus_{j=1}^M h_{ \Lambda_j}\right),
\]
where $h_{\Lambda_j}$ denotes the compression of $h_j$ to $\ell^2(\Lambda_j)$.
By Assumption~\DECtwo(5--6), $V\in\cO_\Lambda$, and the dynamics $\tau_\Lambda$ is generated by
$\delta_\Lambda=\delta_{\fr,\Lambda}+\i[V,\,\cdot\;]$. The reference state
$\omega_\Lambda$ is the quasi-free state on $\cO_\Lambda$ with density
\beq
T_\Lambda=\one_S\oplus\left(\bigoplus_{j=1}^M T_{\Lambda_j,\beta_j\mu_j}\right),\qquad
T_{\Lambda_j, \beta_j\mu_j}=\frac{1}{1 +\e^{\beta_j(h_{ \Lambda_j}-\mu_j\one_{\Lambda_j})}}.
\label{h-sun}
\eeq
Our TDL  scheme is defined by the net $(\cO_\Lambda,\tau_\Lambda,\omega_\Lambda)$.

Other approximation schemes are of course possible, and we will comment on them in Section~\ref{sec-remarks-ebbm}.
\subsection{TDL of 2TMEP}
We start with assumption

\begin{quote}{\bf (FE)} The reservoir's one-fermion Hamiltonians $h_j$, $1\leq j \leq M$, have purely absolutely continuous spectrum.
\label{fe}
\end{quote}
\newcommand{\FE}{{\hyperref[fe]{{\rm (FE)}}}}
This assumption ensures that all reservoir subsystems $(\cO_j,\tau_j,\omega_{\beta_j\mu_j})$ are mixing, and hence
ergodic, and is used only for the verification of Assumption~\TDLThree(2).

\begin{theorem}\label{ebbm-main-thm-1}
Suppose that~\FE{} holds. Then Assumptions~\TDLOne--\TDLSix{} hold. In particular, Proposition~\ref{sl-exam} and
Theorem~\ref{ss-sunday} hold for the {\rm EBBM}.
\end{theorem}

In the process of the proof we will establish regularity properties of the EBBM that are much stronger
than needed for the verification of~\TDLOne--\TDLSix.
\bel \ben
\item $\slim_{\Lambda_j} h_{\Lambda_j}= h_j$.
\item $\slim_\Lambda  h_{\Lambda, \fr}=h_\fr$.
\item $\slim_\Lambda\Delta_{\omega_\Lambda}^{\i\theta}=\Delta_\omega^{\i\theta}$ for all $\theta\in\rr$.
\item For any $A\in\cO_\mathrm{loc}$,
\[
\lim_\Lambda\tau_\Lambda^z(A)=\tau^z(A)
\]
in norm, locally uniformly for $z\in\cc$.
\item For any $A\in\cO$,
\[
\lim_\Lambda\tau_\Lambda^t(A)=\tau^t(A)
\]
in norm, locally uniformly for $t\in\rr$.

\item For all $\Lambda\in\cI$ one has $\sigma_\Lambda=\delta_{\omega_\Lambda}(V)=\sigma$. Moreover,
\[
\lim_\Lambda\varsigma_{\omega_\Lambda}^z(\sigma)=\varsigma_\omega^z(\sigma)
\]
in norm, locally uniformly for $z\in\cc$. In particular, the function $z\mapsto\varsigma_\omega^z(\sigma)$ is entire.
\item For all $s\in\rr$ the map
\[
\i\rr\ni z\mapsto [D\omega_s:D\omega]_z\in\cO
\]
extends to an entire function. Moreover, for $s\in\rr$,
\[
\lim_\Lambda [D\omega_{\Lambda, s}:D\omega_\Lambda]_z=[D\omega_s:D\omega]_z,
\qquad \lim_\Lambda [D\omega_{\Lambda, s}:D\omega_\Lambda]_z^\ast=[D\omega_s:D\omega]_z^\ast,
\]
in norm, locally uniformly for $z\in\cc$.
\een
\label{eret-1}
\eel
\proof {\bf(1)} Since $h_{\Lambda_j}=\one_{\Lambda_j}h\one_{\Lambda_j}$, with $\slim_\Lambda\one_{\Lambda_j}=\one_{R_j}$,
one has the estimate
$$
\|(h_j-h_{\Lambda_j})f\|\le\|(\one_{R_j}-\one_{\Lambda_j})hf\|+\|h(\one_{R_j}-\one_{\Lambda_j})f\|,
$$
which yields the statement. We observe that this implies, in particular, that
$\slim_\Lambda\e^{z h_{\Lambda_j}}=\e^{z h_j}$, locally uniformly for $z\in\cc$.

{\bf(2)} Writing
$$
h_{\Lambda,\fr}-h_\fr=
h_S\oplus\left(\bigoplus_{j=1}^M(h_{\Lambda_j}-h_j)\right),
$$
the result follows directly from~(1). Here again, we note that this implies
$\slim_\Lambda\e^{z h_{\Lambda,\fr}}=\e^{z h_\fr}$, locally uniformly for $z\in\cc$.

{\bf (3)} Let $\ell=-\sum_{j=1}^M\beta_j(h_j-\mu_j)$ and $\ell_\Lambda=-\sum_{j=1}^M\beta_j(h_{\Lambda_j}-\mu_j)$.
By the final remark in the proof of Part~(1), we have $\slim_\Lambda\e^{z\ell_\Lambda}=\e^{z\ell}$,
locally uniformly for $z\in\cc$.
We identify $\Gamma_-(\ell^2(\Lambda))$ with a subspace of $\Gamma_-(\ell^2(G))$ and denote by
$\Gamma_-^{(N)}(\ell^2(G))$ the $N$-particle sector of the latter. Then, the respective
Araki--Wyss representations give that, for any $N,N'\in\nn$, $\theta\in\rr$ and
$\Psi\in\Gamma_-^{(N)}(\ell^2(G))\otimes\Gamma_-^{(N')}(\ell^2(G))$,
\begin{align*}
\Delta_{\omega_\Lambda}^{\i\theta}\Psi
&=\left(\e^{\i\theta\ell_\Lambda}\right)^{\otimes N}
\otimes\left(\e^{-\i\theta\ell_\Lambda}\right)^{\otimes N'}\Psi,\\[4pt]
\Delta_\omega^{\i\theta}\Psi&=\left(\e^{\i\theta\ell}\right)^{\otimes N}
\otimes\left(\e^{-\i\theta\ell}\right)^{\otimes N'}\Psi,
\end{align*}
and the result follows from a simple telescopic expansion of the difference
$\Delta_{\omega_\Lambda}^{\i\theta}\Psi-\Delta_\omega^{\i\theta}\Psi$.

{\bf (4)} Consider first  $A=a^\ast_xa_y$ for some $x,y\in G$.
Since
\[
\begin{split}
\tau_{\Lambda,\fr}^z(A)&=a^\ast(\e^{zh_{\Lambda,\fr}}\delta_x)a(\e^{zh_{\Lambda,\fr}}\delta_y),\\[4pt]
\tau_{\fr}^z(A)&=a^\ast(\e^{zh_\fr}\delta_x)a(\e^{zh_\fr}\delta_y),
\end{split}
\]
one has
\begin{align*}
\|\tau_{\Lambda,\fr}^z(A)-\tau_\fr^z(A)\|&\le\|a^\ast((\e^{zh_{\Lambda,\fr}}-\e^{zh_\fr})\delta_x)\|
\|a(\e^{zh_{\Lambda,\fr}}\delta_y)\|+
\|a^\ast(\e^{zh_\fr}\delta_x)\|\|a((\e^{zh_{\Lambda,\fr}}-\e^{zh_\fr})\delta_y)\|\\[4pt]
&\le\|(\e^{zh_{\Lambda,\fr}}-\e^{zh_\fr})\delta_x\|\|\e^{zh_{\Lambda,\fr}}\delta_y\|
+\|(\e^{zh_{\Lambda,\fr}}-\e^{zh_\fr})\delta_y\|\|\e^{zh_\fr}\delta_x\|,
\end{align*}
and it follows from~(2) that
\beq
\lim_\Lambda\tau_{\Lambda,\fr}^z(A)=\tau_\fr^z(A)
\label{sode-1}
\eeq
in norm, locally uniformly for $z\in\cc$. Since any $A\in\cO_\mathrm{loc}$ is a finite linear
combination of finite products of factors of the form  $a^\ast_xa_y$, \eqref{sode-1}
holds for all $A\in\cO_\mathrm{loc}$ with the required uniformity. Since
$V\in\cO_\mathrm{loc}$, (4) follows from the expansions
\[
 \tau_{\Lambda}^z(A)=\tau_{\Lambda, \fr}^z(A)+
\sum_{n\geq1}z^n\int_{0\le t_ 1\le\cdots\le t_n\le1}\i[\tau_{\Lambda,\fr}^{zt_n}(V),\cdots
,\i[\tau_{\Lambda,\fr}^{z t_1}(V),\tau_{\Lambda,\fr}^{z}(A)]\cdots]\d t_1\cdots\d t_n,
\]
\[
\tau^z(A)=\tau_\fr^z(A)+
\sum_{n\geq1}z^n\int_{0\le t_1\le\cdots\le t_n\leq1}\i[\tau_\fr^{zt_n}(V),\cdots
,\i[\tau_\fr^{zt_1}(V),\tau_\fr^{z}(A)]\cdots]\d t_1\cdots\d t_n,
\]
and a telescopic expansion of their difference.

{\bf (5)} Since $\tau_\Lambda^t$ and $\tau^t$ are isometric for real $t$, the result
follows from Part~(4) and the dense inclusion $\cO_\mathrm{loc}\subset\cO$ by an
elementary $\epsilon/3$ argument.

{\bf (6)} Let $c$ be a finite rank operator on $\ell^2(G)$, with singular value decomposition
$c=\sum_{k=1}^ms_k f_k(g_k,\,\cdot\;)$. Then, $\d\Gamma(c)=\sum_{k=1}^ms_ka^\ast(f_k) a(g_k)\in\cO$,
and one easily checks that
$$
\|\d\Gamma(c)\|\le\|c\|_1=\sum_{k=1}^m s_k\|f_k\|\|g_k\|,
$$
where $\|\,\cdot\,\|_1$ denotes the trace norm. Since finite rank operators are dense in the Banach space
of trace class operators, one has $\d\Gamma(c)\in\cO$ for trace class $c$, with
the same inequality.

By assumption, $v$ is finite rank and $V=W+\d\Gamma(v)$ with $W\in\cO_S$. It follows that
$$
[\d\Gamma(\ell),W]=\sum_{x,y\in G\setminus S}\ell(x,y)[a_x^\ast a_y,W]=0,
$$
which gives
\beq
\varsigma_\omega^\theta(V)
=\e^{\i\theta\d\Gamma(\ell)}V\e^{-\i\theta\d\Gamma(\ell)}
=\e^{\i\theta\d\Gamma(\ell)}W\e^{-\i\theta\d\Gamma(\ell)}
+\d\Gamma\left(\e^{\i\theta\ell}v\e^{-\i\theta\ell}\right)
=W+\d\Gamma\left(\e^{\i\theta\ell}v\e^{-\i\theta\ell}\right),
\label{equ:sigomegaV}
\eeq
and so
$$
\sigma=\delta_\omega(V)
=\frac{\d\ }{\d\theta}\varsigma^\theta_\omega(V)\big|_{\theta=0}=\d\Gamma(\i[\ell,v])
=\sum_{j=1}^M\beta_j\d\Gamma(\phi_j),\qquad\phi_j=\i[v_j,h_j]=\i[h,h_j].
$$
Similarly, recalling that $V\in\cO_\Lambda$ for all $\Lambda\in\cI$, we get
$$
\sigma_\Lambda=\delta_{\omega_\Lambda}(V)=\d\Gamma(\i[\ell_\Lambda,v])
=\sum_{j=1}^M\beta_j\d\Gamma(\phi_{j,\Lambda}),
\qquad\phi_{j,\Lambda}=\i[v_j,h_{j,\Lambda}]=\i[v_j,h_{j}]=\phi_j,
$$
which shows that $\sigma_\Lambda=\sigma$.

For $\Lambda\in\cI$ and $z\in\cc$, we have
$$
\varsigma_{\omega_\Lambda}^z(\sigma)=\d\Gamma\left(\phi_\Lambda(z)\right),\qquad
\varsigma_{\omega}^z(\sigma)=\d\Gamma\left(\phi(z)\right),
$$
where
$$
\phi_\Lambda(z)
=\e^{\i z\ell_\Lambda}\phi\e^{-\i z\ell_\Lambda},
\qquad
\phi(z)
=\e^{\i z\ell}\phi\e^{-\i z\ell},
$$
and $\phi=\sum_{j=1}^M\beta_j\phi_j$ is finite rank. It follows from the initial remark in the proof of Part~(3) that
$$
\lim_\Lambda\phi_\Lambda(z)=\phi(z)
$$
holds in trace norm, locally uniformly for $z\in\cc$, and the result follows from the estimate
$$
\|\varsigma_{\omega_\Lambda}^z(\sigma)-\varsigma_{\omega}^z(\sigma)\|\le\|\d\Gamma(\phi_\Lambda(z)-\phi(z))\|
\le\|\phi_\Lambda(z)-\phi(z)\|_1.
$$

{\bf (7)} We follow the proofs of Propositions~\ref{sd-new} and~\ref{sd-new-new}. Relation~\eqref{equ:sigomegaV} gives
that $\theta\mapsto\varsigma_\omega^\theta(V)$ is entire, from which~\eqref{equ:sigomegagamma} shows that the same
is true of $\theta\mapsto\varsigma_\omega^\theta(\Gamma_{-t})$. It then follows from~\eqref{equ:sigQs}
that~\eqref{st-new} holds for all $z\in\cc$ and gives the first part of the statement.

With~\eqref{st-new-1}, \eqref{sst-new-1}, and~\eqref{equ:sigomegalambdagamma}, to get the second assertion it is
sufficient to prove that
\begin{align*}
\lim_\Lambda\tau_{\Lambda,\fr}^t(\varsigma_{\omega_\Lambda}^\theta(V))&=\tau_\fr^t(\varsigma_\omega^\theta(V)),\\[4pt]
\lim_\Lambda\tau_{\Lambda,\fr}^t(\varsigma_{\omega_\Lambda}^\theta(\sigma))&=\tau_\fr^t(\varsigma_\omega^\theta(\sigma)),
\end{align*}
locally uniformly for $t\in\rr$ and $\theta\in\cc$. This follows from~(5) and~(6). \hfill\qed

\medskip\noindent
{\bf Proof of Theorem~\ref{ebbm-main-thm-1}.}

\TDLOne+\TDLFour{} Follow from Lemma~\ref{eret-1}(7).

\TDLTwo{} Parts~(1--2) are obvious consequences of our setup.
Part~(3) follows from the definition~\eqref{h-sun} of $T_\Lambda$,
Lemma~\ref{eret-1}(1) and $\lim_\Lambda\one_{\Lambda_j}=\one_j$ which ensure that $\slim_\Lambda T_\Lambda=T$.
The general formula for quasi-free states
$$
\omega_T(a^\ast(f_n)\cdots a^\ast(f_1)a(g_1)\cdots a(g_m))=\delta_{nm}\det(\{(g_i,Tf_j)\})
$$
then implies the weak$^\ast$ convergence $\omega_{T_\Lambda}\to\omega_T$. Part~(4) follows from Lemma~\ref{eret-1}(4).
As in the proof of Lemma~\ref{eret-1}(3), we identify $\Gamma_-(\ell^2(\Lambda))$ with a subspace of
$\Gamma_-(\ell^2(G))$. Then the respective Araki--Wyss representations give that Part~(5) holds with
$\Omega_\Lambda=\Omega_\mathrm{AW}=\Omega_\mathrm{f}\otimes\Omega_\mathrm{f}$.\footnote{Note, however, that
$\pi_\Lambda\neq\pi$ on $\cO_\Lambda$ since $T_\Lambda\neq T\big|_{\ell^2(\Lambda)}$.} Part~(6) follows from
Lemma~\ref{eret-1}(7).

\TDLThree{} Part~(1) follows from Lemma~\ref{eret-1}(3) and Formula~\eqref{aw-mod-con} that allow us to take
$J_\Lambda=J$. Regarding Part~(2), setting $R=\cup_{j=1}^MR_j$ and invoking the fermionic exponential
law~\cite[Theorem 3.2]{Baez1992}, we may identify
\[
\Gamma_-(\ell^2(G))=\Gamma_-(\ell^2(S))\otimes\Gamma_-(\ell^2(R)),
\]
and the corresponding Fock vacua
$$
\Omega_\mathrm{f}=\Omega_{\mathrm{f},S}\otimes\Omega_{\mathrm{f},R}.
$$
Considering the operators $\ell$ and $\ell_\Lambda$ defined in the proof of Lemma~\ref{eret-1}(3) as acting on
$\ell^2(R)$, we have
$$
\Delta_\omega^\theta=(I\otimes\e^{\i\theta\d\Gamma(\ell)})\otimes(I\otimes\e^{-\i\theta\d\Gamma(\ell)}),
$$
and
$$
\Delta_{\omega_\Lambda}^\theta
=(I\otimes\e^{\i\theta\d\Gamma(\ell_\Lambda)})\otimes(I\otimes\e^{-\i\theta\d\Gamma(\ell_\Lambda)}).
$$
Since $\d\Gamma(\ell_\Lambda)\Omega_{\mathrm{f},R}=0$, we have
$$
(\Gamma_-(\ell^2(S))\otimes\Omega_{\mathrm{f},R})\otimes(\Gamma_-(\ell^2(S))\otimes\Omega_{\mathrm{f},R})\subseteq
\Ker\log\Delta_{\omega_\Lambda}.
$$
Assumption~\FE{} implies that
\[
\Ker\log\Delta_\omega
=(\Gamma_-(\ell^2(S))\otimes\Omega_{\mathrm{f},R})\otimes(\Gamma_-(\ell^2(S))\otimes\Omega_{\mathrm{f},R}),
\]
which gives Part~(2).

\TDLFive{} Parts~(1--3) follow from our setup, Part~(4) from Lemma~\ref{eret-1}(6).

\TDLSix{} Parts~(1--2) follow from the setup and Parts~(3--4) from Lemma~\ref{eret-1}(2+6). \hfill\qed

\subsection{Remarks}
\label{sec-remarks-ebbm}

{\bf 1.The Spin--Fermion Model.} This open quantum system was studied in the
early seminal works~\cite{Davies1974,Spohn1978b} and has remained one of the
paradigmatic models of quantum statistical mechanics. The transfer operator
techniques of~\cite{Jaksic2002a, Jaksic2010b} are applicable to this model, and
we will return to it in the continuation of this paper~\cite{Benoist2024}. Just
like OQ2S, the Spin--Fermion Model (in the sequel abbreviated SFM) has the
tensor product structure of open quantum systems discussed in Section~1.2
of~\cite{Benoist2023a}. The reservoirs are free Fermi gasses over $(\fh_j,h_j)$
described by $(\CAR(\fh_j),\tau_j,\omega_{\beta_j})$, where $\omega_{\beta_j}$
is a quasi-free state generated by $T_{\beta_j0}$. The interaction of $S$ with
$R_j$ is described by a self-adjoint $V_j$ which is a finite sum of terms
\beq
Q\otimes\varphi_j(f_1)\cdots\varphi_j(f_n),
\label{noi-noi}
\eeq
where $Q$ is a self-adjoint element of $\cO_S$ and $\varphi_j$ denotes the
fermionic field operator in $\CAR(\fh_j)$. If $\fh_j=\ell^2(R_j)$  for some
countably infinite $R_j$, the TDL of the 2TMEP is carried out in the same way as
for the EBMM. The continuous reservoir case $\fh_j=L^2(\rr^d, \d x)$ and
$h_j=-\frac{1}{2m}\Delta$ is of particular importance in connection to  the transfer
operator techniques of~\cite{Jaksic2002a,Jaksic2010b}. The TDL scheme is carried
out by considering subspaces $ L^2([-L,L]^d,\d x)$ and by restricting
$-\frac{1}{2m}\Delta$ to $[-L,L]^d$ with periodic (or Dirichlet, or Neumann...)
boundary condition. To define a finite dimensional TDL scheme, one also
introduces a high energy cut-off $E>0$, and so the scheme is indexed by
$(E,L)_{E>0,L>0}$. The details are the same as
in~\cite[Exercise~6.4]{Jaksic2010b}, and we leave them to an interested reader.
Assumptions~\TDLOne--\TDLSix{} hold if the test functions $f_k$
in~\eqref{noi-noi} are in $\Dom(\e^{\lambda_jh_j})$ for some  $\lambda_j
>\beta_j$.

{\bf 2. The Spin--Boson Model.} This model has the same general structure as the SFM
upon  replacing ${\rm CAR}$ with ${\rm CCR}$. The model cannot be defined on the
$C^\ast$-level and the interacting dynamics is introduced only in the
GNS-representation. For technical reasons due to the unboundedness of the bosonic
fields operators, in~\eqref{noi-noi} one always takes $n\leq 2$. In spite of the
relatively large literature on the Spin--Boson Model, its TDL has rarely been
discussed. Although it is likely that the TDL justification of the 2TMEP for the
Spin--Boson Model can be carried out along similar lines as for the SFM model,
the technical details remain to be worked out.

{\bf 3. Other TDL schemes.} Just like in the OQ2S case~\eqref{new-c}--\eqref{new-c-1},
one can also consider the TDL scheme where
\beq
\omega_{\Lambda_j}=\omega_{\beta_j}\big|_{\cO_{\Lambda_j}}.
\label{new-f}
\eeq
Note that $\omega_{\Lambda_j}$ is quasi-free state on $\cO_{\Lambda_j}$
generated by the compression of $T_{\beta_j\mu_j}$ on the subspace
$\ell^2(\Lambda_j)$, which we denote by $[T_{\beta_j \mu_j}]_{\Lambda_j}$.
Obviously, $[T_{\beta_j \mu_j}]_{\Lambda_j}$ differs from
$T_{\Lambda_j,\beta_j\mu_j}$ given by~\eqref{h-sun}. One then  takes
$$
\bar h_{\Lambda_j}= -\frac{1}{\beta_j} \log ([T_{\beta_j\mu_j}]_{\Lambda_j} (\one_{\Lambda_j}-[T_{\beta_j\mu_j}]_{\Lambda_j})^{-1})  +\mu_j,
$$
so that
\[
[T_{\beta_j \mu_j}]_{\Lambda_j}=\frac{1}{1 +\e^{\beta_j (\bar h_{\Lambda_j}-\mu_j\one_{\Lambda_j})}}.
\]
$\bar h_{\Lambda_j}$ is also  obviously not equal to the compression
$h_{\Lambda_j}$ of $h_j$ to $\ell^2(\Lambda_j)$, but we still have that
$$
\slim_{\Lambda_j}\bar h_{\Lambda_j}=h_j.
$$
This gives that Lemma~\ref{eret-1} holds as formulated with the same proof,
and that the same holds for Theorem~\ref{ebbm-main-thm-1}.

Finally, one can consider a combined TDL scheme where $\omega_{\Lambda_j}$ is
given by~\eqref{new-f} and $h_{\Lambda_j}$ is  the restriction of $h_j$ to
$\Lambda_j$. The proof now requires a slight modification since the groups
$\tau_{\Lambda_j,\fr}$ and $\varsigma_{\omega_{\Lambda_j}}$ do not commute
anymore. However, since $\slim_{\Lambda_j}[h_{\Lambda_j},\bar h_{\Lambda_j}]=0$,
the required changes of  the argument are minor.

{\bf 4. On the role of free reservoirs.} Free reservoirs play a distinguished
role in both experimental and theoretical studies of open quantum systems. The
free reservoir structure brings to the focus interaction processes involving the
small system. In OQ2S, the study of these interaction processes is affected
by the reservoir complexity, and the respective open quantum system remains
poorly understood simply because its reservoirs are poorly understood. From
a mathematical perspective, free Fermi (and Bose) gas reservoirs allow for
a simple implementation of the TDL with respect to its explicitly identifiable
modular structure given by the Araki--Wyss (or Araki--Woods~\cite{Araki1963})
GNS-representation. They also allow for simple criteria ensuring the reservoirs'
ergodicity, a question which remains poorly understood for quantum spin systems.

This being said, the study of OQ2S brings to the forefront some foundational
problems of  quantum statistical mechanics that stem from the pioneering  works
of Araki, Haag, Ruelle, and many others in 1960's and 70's.\footnote{Most of
those developments are summarized in the classical monographs
\cite{Bratteli1987, Bratteli1981}, see also~\cite{Ruelle1969, Israel1979,
Haag1996, Simon1993, Ohya1993, Derezinski2013}.} Although relatively little
progress has been made in the last forty years, these problems remain at the
core of quantum statistical mechanics and are  a challenge for  future
generations.
\bibliographystyle{capalpha}
\bibliography{TDL}
\end{document}